\documentclass[11pt]{article}
\usepackage{axodraw}
\usepackage{epsfig}
\usepackage{amsfonts}
\usepackage{amsmath}
\usepackage{bbm}
\usepackage{cite}
\allowdisplaybreaks[1]

\input pix.sty

\newcommand{\Hc}{\mbox{H.c.}}

\newcommand{\nG}{n_\rmii{$G$}}
\newcommand{\zerob}{0_b}
\newcommand{\zerof}{0_\rmi{\sl f}}
\renewcommand{\mb}{m_b}
\newcommand{\mf}{m_\rmi{\sl f}}
\newcommand{\mphiT}{m_\rmii{$\phi T$}}
\newcommand{\mchiT}{m_\rmii{$\chi T$}}
\newcommand{\E}{\omega}
\newcommand{\mW}{m_\rmii{$W$}}
\newcommand{\mZ}{m_\rmii{$Z$}}

\newcommand{\mWt}{\widetilde m_\rmii{$W$}}

\newcommand{\mh}{m_\rmi{$h$}}
\newcommand{\mt}{m_\rmi{$t$}}
\newcommand{\mG}{m_\rmii{$G$}}
\newcommand{\mht}{\widetilde m_\rmi{$ h$}}
\newcommand{\mGt}{\widetilde m_\rmii{$ G$}}
\newcommand{\mH}{m_\rmii{$H$}}
\newcommand{\mA}{m_\rmii{$A$}}
\newcommand{\mHpm}{m_\rmii{$H_\pm$}}
\newcommand{\mHt}{\widetilde m_\rmii{$ H$}}
\newcommand{\mAt}{\widetilde m_\rmii{$ A$}}
\newcommand{\mHpmt}{\widetilde m_\rmii{$ H_\pm$}}
\newcommand{\mE}{m_\rmii{E2}}
\newcommand{\mC}{m_\rmii{E3}}

\renewcommand{\eq}{eq.~}
\renewcommand{\eqs}{eqs.~}

\renewcommand{\se}{sec.~}
\renewcommand{\ses}{secs.~}
\renewcommand{\fig}{fig.~}

\newcommand{\Tc}{T_{\rm c}}

\newcommand{\gammaE}{\gamma_\rmii{E}}
\newcommand{\rmO}{{\mathcal{O}}}
\newcommand{\bmu}{\bar\mu}

\def\lsi{\raise0.3ex\hbox{$<$\kern-0.75em\raise-1.1ex\hbox{$\sim$}}}
\def\gsi{\raise0.3ex\hbox{$>$\kern-0.75em\raise-1.1ex\hbox{$\sim$}}}
\newcommand{\lsim}{\mathop{\lsi}}
\newcommand{\gsim}{\mathop{\gsi}}

\newcommand{\nF}{n_\rmii{F}}
\newcommand{\nB}{n_\rmii{B}}
 \renewcommand{\nF}[1]{n_\rmii{F{#1}}}
 \renewcommand{\nB}[1]{n_\rmii{B{#1}}}

\newcommand{\rmii}[1]{{\mbox{\tiny\rm{#1}}}}

\newcommand{\re}{\mathop{\mbox{Re}}}

\newcommand{\Tint}[1]{{\hbox{$\sum$}\!\!\!\!\!\!\!\int\,}_{\!\!\!\!\raise-0.9ex\hbox{$\scriptstyle{#1}$}}}
\newcommand{\Tinti}[1]{{{\Sigma}\!\!\!\!\raise0.3ex\hbox{$\int$}_\rmii{${#1}$}}}

\newcommand{\bi}{\begin{itemize}}
\newcommand{\ei}{\end{itemize}}
\newcommand{\hide}[1]{ }
\newcommand{\bsl}[1]{\,\slash\!\!\!\!{#1}\,}

\newcommand{\deltabar}{\delta\!\!\!\raise0.7ex\hbox{--}\,}
\def\TAsc(#1,#2)(#3,#4,#5)%
{\SetWidth{2.0}\CArc(#1,#2)(#3,#4,#5)\SetWidth{1.0}}
\def\Lwidth{3}

\def\TAgl(#1,#2)(#3,#4,#5){\SetWidth{2.0}\PhotonArc(#1,#2)(#3,#4,#5){\Lwidth}%
{6.283 #3 mul 360 div #4 #5 sub #4 #5 sub mul sqrt mul Tdensity mul}%
\SetWidth{1.0}}
\def\TLgl(#1,#2)(#3,#4){\SetWidth{2.0}\Photon(#1,#2)(#3,#4){\Lwidth}
{#1 #3 sub #1 #3 sub mul #2 #4 sub #2 #4 sub mul add sqrt Tdensity mul}%
\SetWidth{1.0}}

\renewcommand{\picc}[1]{\;\parbox[c]{60pt}{\begin{picture}(60,30)(0,0)
\SetWidth{1.0}\SetScale{1.3} #1 \end{picture}}\; }
\def\Lwidth{1.3}

\newcommand{\piB}[1]{\;\parbox[c]{60pt}{\begin{picture}(60,40)(0,0)
\SetWidth{1.0}\SetScale{1.0} #1 \end{picture}}\;}

\def\Deca{\piB{%
 \SetWidth{1.0} 
 \CArc(30,25)(15,0,360)
 \Line(15,25)(45,25)
 \Text(30,0)[c]{{(sss)}}
}}
\def\Decb{\piB{%
 \SetWidth{1.0} 
 \CArc(19.75,25)(10,0,360)
 \CArc(40.25,25)(10,0,360)
 \Text(30,0)[c]{{(ss)}}
}}
\def\Decc{\piB{%
 \SetWidth{1.0} 
 \CArc(30,25)(10,0,360)
 \Line(17,22)(23,28)
 \Line(17,28)(23,22)
 \Text(30,0)[c]{{(s)}}
}}
\def\Decd{\piB{%
 \SetWidth{1.0} 
 \CArc(30,25)(15,0,360)
 \Line(15,25)(45,25)
 \Line(27,22)(33,28)
 \Line(27,28)(33,22)
 \Text(30,0)[c]{{(a)}}
}}
\def\Dece{\piB{%
 \SetWidth{1.0} 
 \CArc(19.75,25)(10,0,360)
 \CArc(40.25,25)(10,0,360)
 \Line(7,22)(13,28)
 \Line(7,28)(13,22)
 \Text(30,0)[c]{{(b)}}
}}
\def\Decf{\piB{%
 \SetWidth{1.0} 
 \CArc(30,25)(10,0,360)
 \Line(17,22)(23,28)
 \Line(17,28)(23,22)
 \Line(37,22)(43,28)
 \Line(37,28)(43,22)
 \Text(30,0)[c]{{(c)}}
}}
\def\Decg{\piB{%
 \SetWidth{1.0} 
 \CArc(19.75,25)(10,0,360)
 \CArc(40.25,25)(10,0,360)
 \Line(7,22)(13,28)
 \Line(7,28)(13,22)
 \Line(47,22)(53,28)
 \Line(47,28)(53,22)
 \Text(30,0)[c]{{(d)}}
}}

\makeatletter \@addtoreset{equation}{section} \makeatother
\renewcommand{\theequation}{\arabic{section}.\arabic{equation}}
\makeatletter
\renewcommand\section{\@startsection {section}{1}{\z@}%
                                   {-5.5ex \@plus -1ex \@minus -.2ex}
                                   {2.3ex \@plus.2ex}%
                                   {\normalfont\large\bfseries}}
\renewcommand\subsection{\@startsection{subsection}{2}{\z@}%
                                     {-3.25ex\@plus -1ex \@minus -.2ex}%
                                     {1.5ex \@plus .2ex}%
                                     {\normalfont\normalsize\bfseries}}
\renewcommand\thesection {\@arabic\c@section}
\renewcommand\thesubsection   {\thesection.\@arabic\c@subsection}
\renewcommand{\@seccntformat}[1]{%
\csname the#1\endcsname.\hspace{1.0em}}
\makeatother

\begin{document}

\flushbottom

\begin{titlepage}

\begin{flushright}
April 2017
\vspace*{1cm}
\end{flushright} 
\begin{centering}

\vfill

{\Large{\bf
 Thermal phase transition 
 with full 2-loop effective potential 
}} 

\vspace{0.8cm}

M.~Laine, 
M.~Meyer, 
G.~Nardini

\vspace{0.8cm}

{\em
AEC, Institute for Theoretical Physics, 
University of Bern, \\ 
Sidlerstrasse 5, CH-3012 Bern, Switzerland\\} 

\vspace*{0.8cm}

\mbox{\bf Abstract}

\end{centering}

\vspace*{0.3cm}
 
\noindent
Theories with extended Higgs sectors constructed in view of
cosmological ramifications (gravitational wave signal, 
baryogenesis, dark matter) are often faced with conflicting
requirements for their couplings; in particular those influencing the
strength of a phase transition may be large. Large couplings
compromise perturbative studies, as well as the high-temperature
expansion that is invoked in dimensionally reduced lattice
investigations. With the example of the inert doublet extension of the
Standard Model (IDM), we show how a resummed 2-loop effective
potential can be computed without a high-$T$ expansion, and
use the result to scrutinize its accuracy. With the exception of $\Tc$,
which is sensitive to contributions from heavy modes, the high-$T$ expansion 
is found to perform well. 2-loop corrections weaken 
the transition in IDM, but they are moderate, whereby 
a strong transition remains an option. 

\vfill

 

\vfill

\end{titlepage}

%
\section{Introduction}

With the upcoming years of the LHC probing the Higgs mechanism, 
and the continued direct, indirect and collider searches for dark matter, 
together with the prospect of LISA probing
gravitational wave backgrounds related to  
particle physics, 
it has become popular to search for
a framework which may play a role in all contexts. Surprisingly, 
the Standard Model supplemented by an additional scalar
field, for instance in the singlet, doublet, triplet, 
or higher representation, 
cannot easily be excluded from these considerations. 
We focus here on the doublet case, simplified further by
an additional Z(2) symmetry, a framework that is 
generally referred to as the Inert Doublet Model (IDM)~\cite{id0,id1,id2}.  

The original interest in the IDM
came largely from the dark matter 
context~\cite{mhgt1,edsjo,mhgt2,agrawal,andreas,multiplet,
arina,dolle,ll1,schannel}, 
which remains a viable option today
(cf.\ e.g.~refs.~\cite{pole,ai1,mk,arhrib,modak,queiroz,ibarra,banerjee} 
and references therein). 
Many theoretical 
(cf.\ e.g.~refs.~\cite{ivanov,chakrabarty,khan,bs,ferreira})
and collider 
(cf.\ e.g.~refs.~\cite{edsjo2,dolle2,mg,aa,krawczyk2,dorsch,belanger,arhrib2,
ikr,blinov2,diaz,mkr2,poulose,kanemura,datta,hashemi,belyaev})
constraints on the model have been considered. Furthermore, 
following early suggestions~\cite{ginzburg,senj,jc1,gch,ck2}, 
a strong phase transition appears possible~\cite{blinov1,pro,basler}. 
However the issue of large couplings emerges, 
for instance in some of the benchmarks of ref.~\cite{pro} certain scalar
couplings attain the magnitude 
$\lambda^{ }_3\simeq 3$ in a normalization in which
the Standard Model Higgs self-coupling is $\lambda_1 \simeq 0.15$.  

There is a clear reason for the need for large couplings
if a strong phase transition is to be present. Without 
any additional particles, the theory has no thermal phase transition
at all (for a review, see ref.~\cite{rev}). If degrees of freedom
are added which are weakly coupled and massive, they can be integrated
out, resulting in the same ``dimensionally reduced''  
effective theory~\cite{dr1,dr2} as for the Standard Model~\cite{generic}, 
and thereby with the same conclusion concerning the 
phase transition. To change the conclusion, we either
need to add new degrees of freedom which are light around the transition
point, or which come with large couplings, so that the effective couplings of 
the low-energy theory change by a significant amount. Light degrees of freedom
could experience a transition of their own and thereby indeed influence 
the dynamics substantially~\cite{twostep1,twostep2}; 
this is an interesting option but will not be considered here, given
that it requires a degree of fine tuning.  
Thereby we are left with large couplings as the remaining avenue. 
It is difficult to exclude the existence
of such couplings phenomenologically,  
given that Higgs physics does not easily avail itself 
to precision inspection and that constraints from fermionic processes 
are largely missing for the inert doublet. Large couplings do imply the 
presence of a nearby Landau pole and, conversely, could originate as 
a low-energy description of some sort of composite dynamics. 

In the context of electroweak baryogenesis,  
a strong phase transition refers to a discontinuity $\Delta v\sim T$, 
where $v$ is a gauge-fixed Higgs expectation value 
($v\simeq 246$~GeV at $T=0$), 
and $T$ is the temperature~\cite{msold}.
In the Higgs phase, gauge boson masses are then of order 
$\mW \sim gv/2 \ll \pi T$, where $g\sim 2/3$ is 
the SU$^{ }_\rmii{L}$(2) gauge coupling. 
In this situation a high-$T$ expansion in $\mW^2 / (\pi T)^2$
works well.\footnote{%
 For bosonic degrees of freedom the high-$T$ expansion also 
 includes non-analytic terms, such as $(\mW^2)^{3/2} / (\pi T)^3$; 
 however any sum-integral only generates a finite number of such terms,  
 associated with Matsubara zero-mode contributions, so that they do not affect
 the convergence of the infinite series.
 } 
The high-$T$ expansion is an ingredient 
for instance in non-perturbative studies based on dimensional 
reduction (cf.\ e.g.\ refs.~\cite{2hdm,mH126,crossover,dono,singlet}
and references therein). 
However, new degrees of freedom which get a mass
through a large coupling $\lambda_3^{1/2} \sim 2$ may become heavy 
in the broken phase,  $\lambda_3^{1/2} v \sim \pi T$. 
Given that the high-$T$ expansion is an asymptotic series, 
it is not clear whether it is numerically 
accurate in such a situation. 

In order to test the convergence
of the high-$T$ expansion, and 
of the perturbative treatment in general, a sufficient loop
order is needed. Here we go to 2-loop level for the effective
potential. Earlier results probing the validity of the 
high-$T$ expansion at 2-loop level, associated
however with large vacuum masses rather than with large couplings, 
can be found in ref.~\cite{old}. Another related investigation, 
albeit restricted to an Abelian theory and without a detailed exposition
of the ``master'' sum-integrals that appear, 
was presented in ref.~\cite{new}. 

The outline of this paper is the following. After defining the IDM and 
the basic observables of our interest
(\se\ref{se:model}), we adopt a simple procedure for implementing the 
thermal resummations that are necessary 
for a consistent computation at finite temperature (\se\ref{se:resum}). 
Intricacies related to renormalization of
the effective potential in the $R^{ }_\xi$ gauge and in 
the presence of resummation are
briefly reiterated (\se\ref{se:renorm}). 
After illustrating our results  
numerically (\se\ref{se:numerics}), 
we collect together our conclusions 
(\se\ref{se:concl}). In a number of appendices, 
the thermal 2-loop ``master'' sum-integrals used 
for representing the effective potential are computed, both without 
and with a high-$T$ expansion 
(appendix~\ref{se:masters}); the 2-loop Feynman diagrams are listed 
in terms of these ``masters'' (appendix~\ref{se:diags}); and 1-loop 
formulae for vacuum renormalization, both as concerns 
the initial values of renormalization group evolution and the
renormalization group evolution itself, are specified 
(appendix~\ref{se:vac}). 

%
\section{Model and observables}
\la{se:model}

In the IDM~\cite{id0,id1,id2}, the Standard Model Higgs
doublet, $\phi$, is supplemented by an additional doublet, $\chi$, which
has the same gauge charges as $\phi$ but in addition displays 
an unbroken global Z(2) symmetry, which forbids Yukawa couplings
to Standard Model fermions. The scalar potential has the form
\ba
 V^\rmii{ }_0 & = &   
 \mu_1^2\, \phi^\dagger \phi^{ }   
 + \mu_2^2\, \chi^\dagger \chi^{ }   
 + \lambda^{ }_1\, ( \phi^\dagger \phi^{ } )^2 
 + \lambda^{ }_2\, ( \chi^\dagger \chi^{ } )^2 \nn  
 & + &
   \lambda^{ }_3\, \phi^\dagger \phi^{ }\, \chi^\dagger \chi^{ }
 + \lambda^{ }_4\, \phi^\dagger \chi^{ }\, \chi^\dagger \phi^{ }
 + \Bigl\{ \frac{\lambda^{ }_5}{2}\, (\phi^\dagger \chi^{ })^2 + \Hc\Bigr\}
 \;. \la{V} 
\ea
A global phase rotation permits for us to choose $\lambda^{ }_5$ purely real
and {\em negative}. 
Several extensions of the IDM have
also been proposed, with additional scalars and additional 
gauge symmetries; many lead to fascinating phenomenology but
for brevity we restrict ourselves to the simplest case here, since 
this is sufficient for our methodological considerations. 

We are interested in the behaviour of the model at finite temperature. 
Simple thermodynamic characteristics of a phase transition are its critical 
temperature ($\Tc$) and latent heat~($L$). 
The discontinuity of the Higgs condensate (cf.\ e.g.\ ref.~\cite{bumuwi}), 
\be
 \frac{v_\rmi{phys}^2}{2} \,\equiv\, 
 {Z}^{ }_{\mu_1^2}\, \Delta  \langle \phi^\dagger \phi^{ } \rangle
 \;, 
\ee
where ${Z}^{ }_{\mu_1^2}$ is the renormalization factor related
to the bare mass parameter $\mu_1^2$, 
is a gauge-independent
but scale-dependent
characteristic of the transition. We choose the fixed
$\msbar$ renormalization scale
$\bmu = \mZ^{ }$ for its definition. 
Denoting by $f \equiv F/V$ the free energy density
and by $\Delta f$ its discontinuity across the transition
(with $\Delta f = 0$ precisely at $\Tc$),
we can equivalently 
write $ v_\rmi{phys}^2/2 = \partial \Delta f / \partial \mu_1^2(\mZ)$. 
The importance of $v^{ }_\rmi{phys}$ stems from the fact
that it is strongly correlated with the rate of anomalous baryon 
number violation~\cite{sphaleron1}.

Other important characteristics of the transition are
its surface tension at $\Tc$, 
and the bubble nucleation rate in the whole metastability range. 
Determining these necessitates, however, the study of 
inhomogeneous configurations, which is 
a notoriously hard problem (cf.\ e.g.\ ref.~\cite{gako})
and not addressed here. However our conclusions do support 
a low-energy effective theory approach, which can 
subsequently also be 
applied to this problem~\cite{bubble}.

As a tool for computing $\Tc$, $L$ and $v^{ }_\rmi{phys}$ we employ
the effective potential. At its minima, the effective potential equals
the free energy density, $f = V(v^{ }_\rmi{min})$, up to an overall
constant which drops out in $\Delta f \equiv V(v^{ }_\rmi{min}) - V(0)$. 
The effective potential is defined through a shift
of the neutral Higgs component by a constant, $v$, so that 
the Higgs doublets can be written as
\be
 \phi^{ }_\rmii{R} = \frac{1}{\sqrt{2}}
 \left( 
  \begin{array}{c} G^{ }_2 + i G^{} _1 \\ 
                   v + h - i G^{ }_3
  \end{array}
 \right)
 \;, \quad
 \chi^{ }_\rmii{R} = \frac{1}{\sqrt{2}}
 \left( 
  \begin{array}{c} H^{ }_2 + i H^{} _1 \\ 
                   H^{ }_0 - i H^{ }_3
  \end{array}
 \right)
 \;. \la{shift}
\ee
Here $h$ represents the physical Higgs boson, and 
$H \equiv H^{ }_0, A \equiv H^{ }_3$ as well as 
$H^{ }_{\pm} = (H^{ }_1 \pm i H^{} _2)/\sqrt{2}$ are 
the new scalar degrees of freedom. The Z(2) symmetry
associated with $\chi$ is assumed to be unbroken, 
and we check this assumption {\it a posteriori} (see below). 
The meaning of $v$ 
depends on the gauge choice and also on the renormalization
factors ${Z}^{ }_\phi$, ${Z}^{ }_v$ (see below).
Nevertheless, as has been demonstrated within the high-$T$
expansion both for covariant~\cite{covariant} and 
$R^{ }_\xi$~\cite{Rxi} gauges, gauge independent observables
can be obtained from $V(v)$, in particular 
$\Delta f = V(v^{ }_\rmi{min}) - V(0)$, which 
in turn fixes $\Tc$, $L$, and  
$v^{ }_\rmi{phys}$ as discussed above. 

%
\section{Resummation}
\la{se:resum}

When we are addressing the regime $v\lsim T$, then the masses generated
by the Higgs mechanism, $\mW \sim g v/2$, are of a similar magnitude or
smaller than thermal ``Debye masses'', $\mE \sim gT$. Therefore
thermal masses play an important role. For the 
Standard Model, all thermal masses were determined in ref.~\cite{meg}, 
and a way to incorporate them at 2-loop level was worked out in 
ref.~\cite{ae}. However, even though theoretically consistent, 
the procedure of ref.~\cite{ae} is simple only in 
a setting in which a high-$T$ expansion is valid: 
technically it amounts to carrying out a resummation only for 
Matsubara zero modes, for which it is strictly necessary. 
In our more general setting, in which some degrees of freedom
may become heavy in the broken phase, a split-up into 
zero and non-zero Matsubara modes is cumbersome. Therefore, 
we propose to implement a {\em resummation for all modes}.\footnote{%
 Analogous procedures have been pursued in other contexts, 
 cf.\ e.g.\ refs.~\cite{scr0,scr05,scr1,scr2,scr3} and references therein. 
 } 
Of course, there is a price to pay for this 
``simplification'', discussed at the end
of this section and in \se\ref{ss:illustration}.

The set of fields for which a resummation is needed comprises
the scalar fields ($\phi,\chi$) as well as, in covariant and 
$R^{ }_\xi$ gauges, the temporal components of the gauge fields. 
Consider $\phi$ as an example. The original (imaginary-time) 
Lagrangian can be written as 
\ba
 L^{ }_\rmii{B}(\phi^{ }_\rmii{B})
 & = & L(\phi) + \delta L(\phi) \nn 
 & = & 
  L(\phi) + \delta \mphiT^2\, \phi^\dagger\phi + \delta L(\phi) 
 - \delta \mphiT^2\, \phi^\dagger\phi
 \;. \la{resum}
\ea
Here $\phi^{ }_\rmii{B}$ denotes a bare and $\phi \equiv \phi^{ }_\rmii{R}$ 
a renormalized field, 
and $\delta L$ contains the vacuum counterterms. Resummation can now
be implemented by incorporating
$ +\, \delta \mphiT^2\, \phi^\dagger\phi $
on par with vacuum masses in the propagators, 
whereas the part 
$\delta L(\phi) - \delta \mphiT^2\, \phi^\dagger\phi$ 
is treated as a ``counterterm''. If we choose 
$\delta \mphiT^2$ properly, i.e.\ as the thermal mass
generated for the Matsubara zero modes, 
then this procedure is equivalent
to the approach of ref.~\cite{ae}, up to corrections that are
of higher order in couplings than the computation at hand. 

Even though the idea just introduced is simple, the devil lies in the details,
particularly in the precise choice of $\delta \mphiT^2$. 
At high temperatures, the parametric form is 
$
 \delta \mphiT^2 \sim g^2 T^2 + \rmO(g^4)
$.
The factor $T^2$ originates
from integrating out the non-zero Matsubara modes, so that 
$
 \delta \mphiT^2 \sim g^2 I^{ }_{n\neq 0}(m) + \rmO(g^4)
$, 
where 
$I^{ }_{n\neq 0}(m) \equiv \Tinti{P}\!\!' \frac{1}{P^2 + m^2}$ is 
a tadpole integral omitting a Matsubara zero mode.
Following a frequent convention,
we make a choice in the following that 
$
 \delta \mphiT^2 \sim g^2 I^{ }(0) 
 \equiv g^2 \Tinti{P}\! \frac{1}{P^2}
$, 
thereby 
omitting corrections of $\rmO(g^4 T^2)$ and $\rmO(g^2 m^2)$. 
Both approximations
can be systematically lifted within the framework of dimensionally reduced
theories, whereas within the approach of
\eq\nr{resum} there is no unambiguous way to do this.  
We choose the simple procedure because
it is sufficient for addressing the main goals of our study, namely
the convergence of the high-$T$ expansion and the magnitude 
of 2-loop corrections. However, it should be acknowledged that this 
approximation is numerically questionable for the IDM: large scalar couplings
imply that corrections of $\rmO(\lambda_3^2 T^2)$ can be significant, and 
a large mass parameter $\mu_2^2$ implies that mass-dependent 
corrections of $\rmO(\lambda^{ }_3 \mu_2^2)$ 
should be included. The omission of these corrections 
leads to specific problems, discussed below.  

In dimensional regularization, where the space-time
dimension is $D = 4 - 2\epsilon$, the sum-integral 
$I(0)$ contains terms of $\rmO(\epsilon)$, cf.\ \eq\nr{I1b}. 
In loop diagrams $I(0)$ can be multiplied by $1/\epsilon$, 
and therefore $\rmO(\epsilon)$ contributions can give
finite results. The thermal
masses including these pieces are listed in 
\eqs\nr{mphiT}--\nr{mC} below.  

A formal crosscheck on the consistency of the resummation 
carried out is that the so-called  ``linear terms'' 
cancel in the 2-loop result within 
the high-$T$ expansion~\cite{linear}. 
Such terms have the form $\sim g^2 I(0) I^{ }_{n=0}(m)$, where the 
Matsubara zero-mode part evaluates to 
$
 I^{ }_{n = 0}(m) \equiv T \int_\vec{p} \frac{1}{p^2 + m^2} 
 = -\frac{m T}{4\pi} [ 1  + \rmO(\epsilon) ] 
$.
We have analytically verified the cancellation of linear terms to all orders 
in $\epsilon$ within our resummation. However, as alluded to above, 
our simple resummation does not properly capture the infrared structure 
of the 2-loop potential in the IDM, in which substantial corrections of 
$\rmO(\lambda^{ }_3 \mu_2^2)$ to the
effective Higgs mass parameter can appear. This implies that the 
cancellation of linear terms is incomplete beyond the formal 
high-$T$ limit: 
a remainder $\sim \lambda^{ }_3 [I(\mu^{ }_2) - I(0)] I^{ }_{n=0}(m)$
is left over. In cases with $\mu^{ }_2 \gsim T$ such terms become visible 
at small $v/T$ (cf.\ \se\ref{ss:num}).

%
\section{Gauge fixing and renormalization}
\la{se:renorm}

%
\subsection{Gauge fixing}

Perturbative computations in gauge theories require gauge fixing, 
even though physical observables are independent of it. 
For simplicity we employ the Feynman $R^{ }_\xi$ gauge in our analysis. 
We also omit the hypercharge U$^{ }_\rmii{Y}(1)$ coupling $g^{ }_1$, 
whose $\rmO(1\%)$ influence is an order of magnitude smaller
than our uncertainties, and
denote the weak SU$^{ }_\rmii{L}$(2) coupling by $g \equiv g^{ }_2$.
Then the gauge fixing and Faddeev-Popov terms read
\be
 L^{ }_\rmi{gauge fixing} = 
 \frac{1}{2 \xi^{ }_{ }} 
 \sum_{a = 1}^3
 \Bigl( 
   \partial^{ }_\mu A^a_{\mu} - 
   \frac{ \xi^{ }_{ } g^{ }_{ } v^{ }_{ }}{2}
   \, G^{ }_{a} 
 \Bigr)^2
 + \frac{ \xi^{ }_{ } g^2_{ } v^{ }_{ } }{4}
 \Bigl(
  \bar{c}^a_\rmii{$A$}\, h^{ }_{ }\, c^a_\rmii{$A$} + 
  \epsilon^{abc} \bar{c}^a_\rmii{$A$}\, G^{ }_{b}\, c^c_\rmii{$A$} 
 \Bigr) + \ldots
 \;, \la{gauge_fix}
\ee
where only terms coupling to scalar degrees of freedom 
have been shown; 
$G^{ }_a$ are Goldstone modes from \eq\nr{shift}; 
and $ c^{a}_\rmii{$A$},  \bar{c}^a_\rmii{$A$}$ are 
SU$^{ }_\rmii{L}$(2) ghost fields. 

In $R^{ }_\xi$ gauges the parameter $v$ has two different origins: 
it appears as a ``background field'' in the gauge fixing term
in \eq\nr{gauge_fix}, and
it originates from a shift of the Higgs field according to 
\eq\nr{shift}. For a proper renormalization of gauge-dependent
quantities, these two fields need to be kept track of and renormalized
separately (cf.\ refs.~\cite{rxi1,rxi2} and references therein). 
The renormalization factor related to the background field is 
denoted by $Z^{ }_v$: 
$
 v^2_\rmii{B} = v^2 (1 + \delta Z^{ }_v)
$.

%
\subsection{Vacuum counterterms}
\la{ss:cts}

The bare couplings are expressed as 
$
 \mu_{i\rmii{B}}^2 = \mu_i^2 \,
 \bigl( 1 + \delta Z^{ }_{\mu^{2}_i} \bigr)
$, 
$
 \lambda^{ }_{i\rmii{B}} = \lambda^{ }_i \, 
 \bigl( 1 + \delta Z^{ }_{\lambda^{ }_i} \bigr)
$,
$
 g^2_\rmii{B} = g^2 \, 
 \bigl(1 + \delta Z^{ }_{g^2} \bigr)
$,
$
 h^2_{t\rmii{B}} = h_t^2 \, 
 \bigl(1 + \delta Z^{ }_{h^{2}_t}\bigr) 
$, 
where $h^{ }_t$ is the top Yukawa coupling. 
Because we compute the effective potential as a function 
of $v$ and because resummation
treats $A^a_0$ and $A^a_i$ separately, 
we also need to renormalize certain unphysical objects, 
namely wave functions and the gauge fixing parameter: 
\ba
 && 
 \phi^\dagger_\rmii{B} \phi^{ }_\rmii{B} = 
 \phi^\dagger \phi \, 
 \bigl( 1 + \delta Z^{ }_\phi \bigr) 
 \;, \quad
 A^a_{\mu\rmii{B}}  A^a_{\nu\rmii{B}} = 
 A^a_{\mu} A^a_{\nu } \, 
 \bigl( 1 + \delta Z^{ }_A \bigr) 
 \;, \quad
 \xi^{ }_\rmii{B} = \xi \, 
 \bigl( 1 + \delta Z^{ }_\xi \bigr)
 \;. \la{ren_fields}
\ea
The renormalized gauge parameter is set to $\xi = 1$. 
After the shift
of the Higgs vacuum expectation value according to \eq\nr{shift}, various
counterterms are generated. For instance 
$\delta L$ from \eq\nr{resum} becomes 
\ba
 \delta L(\phi) & = & \fr12 h  
 \bigl(- \partial^2 \delta Z^{ }_\phi + \delta \mh^2 \bigr) h
 + \fr12 G_a^{ } 
 \bigl(- \partial^2 \delta Z^{ }_\phi + \delta \mG^2 \bigr) G_a^{ } 
 + \fr14 \delta \lambda^{ }_1 h^4 
 + \ldots 
 \;, \\ 
 \delta \mh^2 & = & 
 \delta \mu_1^2 
 + 3 \,
 \delta \lambda^{ }_1 v^2  
 \;, \quad 
 \delta \mu_1^2 \; = \; 
 \mu_1^2 \, \bigl( \delta Z^{ }_\phi + \delta Z^{ }_{\mu^{2}_1} \bigr)
 \;, \quad 
 \delta \lambda^{ }_1 \; = \; 
 \lambda^{ }_1  \, 
  \bigl( 2 \delta Z^{ }_\phi + \delta Z^{ }_{\lambda^{ }_1} \bigr)
 \;, \la{ct_h} \hspace*{5mm} \\ 
 \delta \mG^2 & = & 
 \mu_1^2 \, \bigl( \delta Z^{ }_\phi + \delta Z^{ }_{\mu^{2}_1} \bigr)
 +  
 \lambda^{ }_1 v^2 \, 
 \bigl( 2 \delta Z^{ }_\phi + \delta Z^{ }_{\lambda^{ }_1} \bigr)
 +  
 \mW^2\, (\delta Z^{ }_\phi 
 + \delta Z^{ }_v + \delta Z^{ }_{g^2} + \delta Z^{ }_\xi)
 \;. \hspace*{7mm}
\ea
Physical on-shell Green's functions, such as those listed
in appendix~\ref{ss:couplingsmZ} for purposes of vacuum 
renormalization, are not affected by the  
field renormalization constants~\cite{sirlin}. 

%
\subsection{Thermal masses}

%
\begin{table}[t]

\begin{center}
\begin{tabular}{llll}
 \hline \\[-3mm]
    field \hspace*{9mm}   &   vacuum mass  \hspace*{34mm}  &
    thermal mass  \hspace*{16mm}  &   degeneracy   \\ 
 \hline
    $h$      & $\mh^2 = \mu_1^2 + 3 \lambda^{ }_1 v^2$ &
             $\mht^2 = \mh^2 + \delta \mphiT^2$  & 1 \\  
 $G^{ }_{ }$   & $\mG^2 = \mu_1^2 + \lambda^{ }_1 v^2 + \mW^2$  &
             $\mGt^2 = \mG^2 + \delta \mphiT^2 $  & 3 \\
    $H$      & $ \mH^2 = \mu_2^2 +
               \frac{1}{2}(\lambda^{ }_3 + \lambda^{ }_4 + \lambda^{ }_5)v^2$
             & $\mHt^2 = \mH^2 + \delta \mchiT^2 $ & 1 \\ 
    $A$      & $ \mA^2 = \mu_2^2 +
               \frac{1}{2}(\lambda^{ }_3 + \lambda^{ }_4 - \lambda^{ }_5)v^2$
             & $\mAt^2 = \mA^2 + \delta \mchiT^2 $ &  1 \\ 
$H^{ }_{\pm}$ & $ \mHpm^2 = \mu_2^2 +
               \frac{1}{2}\lambda^{ }_3v^2$
             & $\mHpmt^2 = \mHpm^2 + \delta \mchiT^2 $ & 2  \\ 
   $ A^{ }_i $ & $\mW^2 = \frac{1}{4}g^2 v^2 $ &
             $\mW^2$ & $3(D-1)$ \\ 
   $ A^{ }_0 $ & $\mW^2$  & $\mWt^2 = \mW^2 + \mE^2 $ & 3 \\ 
   $ c^{ }_\rmii{$A$}, \bar{c}^{ }_\rmii{$A$} $ 
     & $\mW^2$  & $\mW^2$ & $-\,6$ \\ 
   $ C^{ }_i $ & $0$ &
             $0$ & $8(D-1)$ \\ 
   $ C^{ }_0 $ & $0$  & $\mC^2 $ & 8 \\ 
   $ c^{ }_\rmii{$C$}, \bar{c}^{ }_\rmii{$C$} $ & $0$ &
             $0$ & $-\,16$ \\  
   $ t $     & $\mt^2 = \frac{1}{2}h_t^2 v^2$ & $\mt^2$ & 12 \\
 \hline 
\end{tabular} 
\end{center}

\vspace*{3mm}

\caption[a]{\small
 Tree-level masses squared in the Feynman $R^{ }_\xi$ gauge
 (cf.\ \se\ref{se:renorm}). The thermal mass corrections
 $ \delta \mphiT^2 $, $ \delta \mchiT^2 $, 
 $ \mE^2 $ and $ \mC^2$ are given in  
 \eqs\nr{mphiT}--\nr{mC}.
 By $D=4-2\epsilon$ we denote the dimensionality of spacetime. 
 The fields $A^{ }_\mu$ and $C^{ }_\mu$ correspond to the 
 gauge groups SU$^{ }_\rmii{L}$(2) and SU$^{ }_{ }$(3), 
 respectively, with $c^{ }_\rmii{$A$},
 \bar{c}^{ }_\rmii{$A$}$, 
 $c^{ }_\rmii{$C$},
\bar{c}^{ }_\rmii{$C$}$ being the Faddeev-Popov ghosts.
 }
\label{table:masses}
\end{table}
%

With the setup introduced, 
the thermal mass corrections for \eq\nr{resum} 
and table~\ref{table:masses} read
\ba
 \delta \mphiT^2 & = & 
 \biggl[ 
   6 \lambda^{ }_1 + 2 \lambda^{ }_3 + \lambda^{ }_4 + 
   \frac{3(D-1)g_2^2}{4}
 \biggr]\, I(\zerob) - 6 h_t^2 I(\zerof)
 \la{mphiT} \\ & = & 
 \biggl( \frac{3 g_2^2}{16} + \frac{h_t^2}{4}
 + \frac{6 \lambda^{ }_1 + 2 \lambda^{ }_3 + \lambda^{ }_4}{12} \biggr)\, T^2
 + \rmO(\epsilon\, T^2)
 \;, \nn
 \delta \mchiT^2 & = & 
 \biggl[ 
   6 \lambda^{ }_2 + 2 \lambda^{ }_3 + \lambda^{ }_4 + 
   \frac{3(D-1)g_2^2}{4}
 \biggr]\, I(\zerob)  
 \la{mchiT} \\ & = & 
 \biggl(  \frac{3 g_2^2}{16} 
 + \frac{6 \lambda^{ }_2 + 2 \lambda^{ }_3 + \lambda^{ }_4}{12} \biggr)\, T^2
 + \rmO(\epsilon\, T^2)
 \;, \nn
 \mE^2 & = &
 g_2^2 (D-2) \, \Bigl[ 
  2 (D-1) I(\zerob) - 4 \nG I(\zerof) 
 \Bigr]
 \la{mE} \\
 & = & 
 \Bigl( 1 + \frac{\nG}{3} \Bigr) g_2^2 T^2
 + \rmO(\epsilon\, T^2)
 \;,  \nn 
 \mC^2 & = &
 g_3^2 (D-2) \, \Bigl[ 
  3 (D-2) I(\zerob) - 4 \nG I(\zerof) 
 \Bigr] 
 \la{mC} \\
 & = & 
 \Bigl( 1 + \frac{\nG}{3} \Bigr) g_3^2 T^2
 + \rmO(\epsilon\, T^2)
 \;, \nonumber  \hspace*{6mm}
\ea
where the function $I$ is defined in \eq\nr{I}. 
We have adopted a notation in which
$\zerob$ and $\zerof$ denote vanishing masses carried by bosons
and fermions, respectively; 
and $\nG \equiv 3$ denotes the number of generations.  
The resulting tree-level
mass spectrum is listed in table~\ref{table:masses}.

As alluded to in the paragraphs below \eq\nr{resum}, the leading-order
``massless'' resummation of \eqs\nr{mphiT}--\nr{mC} is not sufficient
for a precise determination of $\Tc$: 
the effective Higgs mass parameter gets large corrections of 
$\rmO(\lambda^{ }_3\mu_2^2, \lambda_3^2 T^2)$ which are not
properly accounted for. These corrections could 
be systematically included in a dimensionally reduced 
investigation~\cite{generic,old,singlet}, whose principal accuracy
our study aims to justify. 

%
\subsection{Illustration of cancellation of divergences}
\la{ss:illustration}

The resummation introduced in \eq\nr{resum} modifies the divergence
structure of the theory at any given loop order. Even though the changes
are of higher order than the computation carried out, 
this leads to 
divergences which look worrisome at first sight. 
We illustrate this 
with the help of a single-component scalar theory, 
\be
 V^{ }_0 = \frac{ \mu_1^2\, h^2 }{2}
 \, \Bigl( 1 + \delta Z^{ }_\phi + \delta Z^{ }_{\mu_1^2} \Bigr) 
 + \frac{ \lambda^{ }_1 h^4 }{4} 
 \, \Bigl( 1 + 2 \delta Z^{ }_\phi 
  + \delta Z^{ }_{\lambda^{ }_1} \Bigr) 
 \;. \la{V0_exp}
\ee
At 1-loop level the counterterms read 
$ 
 \delta Z^{ }_\phi = 0
$, 
$
 \delta Z^{ }_{\mu_1^2} = 3 \lambda^{ }_1 / (16\pi^2\epsilon)
$
and
$
 \delta Z^{ }_{\lambda^{ }_1} = 9 \lambda^{ }_1 / (16\pi^2\epsilon)
$.
The thermal mass correction is $\delta \mphiT^2 = 3 \lambda^{ }_1 I(0)$, and 
we denote $\mht^2 = \mu_1^2 + 3 \lambda^{ }_1 v^2 + \delta \mphiT^2$.

The 1-loop effective potential is given by the function $J$ defined
in \eq\nr{J}: $V_1^{ }= J(\mht)$. Writing $V^{ }_{1} = 
\sum_{n=-1}^{\infty} V_1^{(n)}\epsilon^n$, 
let us consider the divergent part, 
given by \eq\nr{Jdiv}, {\it viz.}\ 
\ba
 \frac{ V^{(-1)}_1}
 {\epsilon} & = & 
 -\frac{\mht^4}{64\pi^2 \epsilon} 
 \nn & = &
 -\frac{ 6 \lambda^{ }_1 \mu_1^2 v^2
  + 9 \lambda_1^2 v^4
  + 2  \delta \mphiT^2\, [\mu_1^2 + 3 \lambda^{ }_1 v^2 ] 
  }{64\pi^2\epsilon}
 \; + \; 
 \mbox{($v$-independent)}
 \;. \hspace*{5mm} \la{V1_exp}
\ea
The $T$-independent divergences $\propto \mu_1^2 v^2, v^4$
are cancelled by the
counterterms in \eq\nr{V0_exp}. In contrast, the $T$-dependent divergence
$\propto \delta \mphiT^2$ is only cancelled by a part of $V^{ }_2$, 
as we show below.  This is the peculiarity related to thermal divergences
within the resummation we have adopted: 
whereas vacuum divergences are cancelled by counterterms originating from 
{\em lower-order} diagrams ($V^{ }_0$), 
thermal divergences are cancelled by the appearance
of $\delta \mphiT^2$ within {\em higher-order} contributions
($V^{ }_2$). 

%
\begin{figure}[t]
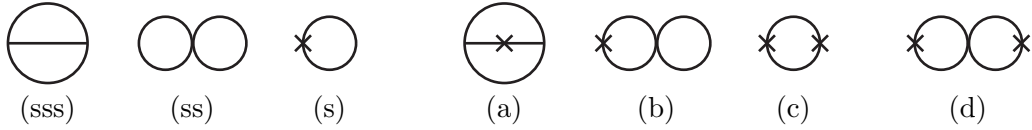


\begin{eqnarray*}
&& 
 \hspace*{-1cm}
 \Deca
 \hspace*{-0.4cm}
 \Decb
 \hspace*{-0.5cm}
 \Decc
 \hspace*{-0.0cm}
 \Decd
 \hspace*{-0.3cm}
 \Dece
 \hspace*{-0.5cm}
 \Decf
 \hspace*{-0.0cm}
 \Decg
\end{eqnarray*}

\caption[a]{\small 
 Topologies of the diagrams discussed in \se\ref{ss:illustration}. 
 The cross denotes a contribution from counterterms including
 the  ``thermal mass'', 
 $-\delta m^2_{\phi\rmii{$T$}}\, \phi^\dagger\phi$
 in \eq\nr{resum}, 
 which is counted on par with a vertex in the resummed computation. 
 Therefore, the graphs (sss), (ss) and (s) are of ``2-loop order''; 
 the graphs (a), (b) and (c) of ``3-loop order''; and the graph
 (d) of ``4-loop order''. 
} 
\la{fig:graphs}
\end{figure}
%

There are three diagrams contributing to $V^{ }_2$
(cf.\ \fig\ref{fig:graphs}): a ``sunset'' diagram
containing three propagators, denoted by (sss); a ``figure-8'' 
diagram containing two propagators, denoted by (ss); and a 
counterterm diagram containing one propagator, denoted by (s). Making
use of the notation of appendix~A, the expressions read
\ba
  V^{ }_2 & = & \mbox{(sss)} + \mbox{(ss)} + \mbox{(s)}
 \;, \\[2mm] 
 \mbox{(sss)} & = & 
 -3 v^2 \lambda_1^2 \, H(\mht,\mht,\mht) \;, \la{sss_exp} \\[2mm] 
 \mbox{(ss)} & = & 
 \frac{3\lambda^{ }_1  I(\mht) I(\mht)}{4} \;, \la{ss_exp} \\ 
  \mbox{(s)} & = & 
  \frac{
  \mu_1^2 \, \delta Z^{ }_{\mu^{2}_1}
 + 3 
 \lambda^{ }_1 v^2 \,  \delta Z^{ }_{\lambda^{ }_1} 
  - \delta \mphiT^2 }{2} \, I(\mht) 
  \;. \la{s_exp}
\ea

The sum of \eqs\nr{sss_exp}--\nr{s_exp} contains divergences of 
orders $1/\epsilon^2$ and $1/\epsilon$, cf.\ \eqs\nr{Idiv}, \nr{Hdiv2}
and \nr{Hdiv1}. 
Writing $V^{ }_2  = \sum_{n=-2}^{\infty} V^{(n)}_2 \epsilon^n$, 
the $1/\epsilon^2$ divergences sum up to 
\be
 \frac{V_2^{(-2)}}{\epsilon^2}
 = \frac{3 \lambda^{ }_1
 [\delta \mphiT^2 + \mu_1^2 + 3 \lambda{ }_1 v^2 ]
 [\delta \mphiT^2 - \mu_1^2 - 9 \lambda{ }_1 v^2 ]
 }{4(4\pi)^4\epsilon^2}
 \;. 
\ee
The $T$-independent parts $\propto \mu_1^2 v^2, v^4$
can be taken care of by 2-loop contributions to
$\delta Z^{ }_{\phi}$,  
$\delta Z^{ }_{\mu_1^2}$ and $\delta Z^{ }_{\lambda^{ }_1}$
in \eq\nr{V0_exp}. 
However, a $T$ and $v$-dependent part 
$ \sim \lambda_1^2 v^2 \delta \mphiT^2 / \epsilon^2$
remains. 
This is only cancelled by diagrams in which the insertion
$\delta \mphiT^2$ appears inside 2-loop topologies, which 
are counted on par with diagrams of 3-loop order 
(diagrams (a) and (b) in \fig\ref{fig:graphs}). 
There is also a $T$-dependent but $v$-independent term 
$ \sim \lambda_1^{ } \delta \mphiT^4 / \epsilon^2$ which 
is cancelled by 2-loop topologies containing two 
insertions of $\delta \mphiT^2$, 
a contribution which is counted on par with diagrams of 4-loop order
(diagram (d) in \fig\ref{fig:graphs}). 

As far as the divergences of order $1/\epsilon$ go, the term proportional
to $\delta \mphiT^2$ in \eq\nr{s_exp} exactly cancels against the 1-loop
contribution in \eq\nr{V1_exp}, apart from a $v$-independent divergence 
$\sim \delta \mphiT^4 /\epsilon$. This gets cancelled
by a 1-loop topology dressed by two appearances of $\delta \mphiT^2$, 
which is counted on par with diagrams of 3-loop order
(diagram (c) in \fig\ref{fig:graphs}). 

A non-trivial cancellation is observed by considering 
$1/\epsilon$-divergences proportional to the non-analytic 
structure 
$
 - {\mht^2} / {(4\pi)^2} \ln\bigl( {\bmu^2} / {\mht^2} \bigr) + 
 I^{(0)}_\rmii{$T$}(\mht)
$, originating from 
$I^2(\mht)$ and \linebreak $H(\mht,\mht,\mht)$ 
as shown by \eqs\nr{I0} and \nr{Hdiv1}, respectively. 
We find that vacuum parts $\propto \lambda^{ }_1\mu_1^2,\lambda_1^2v^2$ 
cancel from the coefficient 
of this divergence, as is required by renormalizability. 
The remainder reads
\be
 \left. \frac{V_2^{(-1)}}{\epsilon} \right|^{ }_\rmi{non-analytic}
 \; = \;  
 - \frac{3 \lambda^{ }_1\delta \mphiT^2}{2(4\pi)^2\epsilon}
 \, 
 \biggl[ 
 - \frac{\mht^2}{(4\pi)^2} \ln\biggl( \frac{\bmu^2}{\mht^2} \biggr) + 
 I^{(0)}_\rmii{$T$}(\mht)
 \biggr]
 \;. 
 \la{non-analytic}
\ee
Again this is only cancelled by diagrams in which the insertion
$\delta \mphiT^2$ appears inside 2-loop topologies, which 
are counted on par with 3-loop graphs
(diagrams (a) and (b) in \fig\ref{fig:graphs}). 

To summarize, one price we pay for the resummation  
introduced in \se\ref{se:resum} is that ultraviolet divergences
do not cancel order by order in our power counting, in which 
the last term of \eq\nr{resum} is treated as an insertion. 
Instead thermal 
divergences cancel once all insertions contributing to a 
given order in couplings have been accounted for. This ``drawback''
is compensated for by the fact that infrared sensitive terms
get resummed to all orders. 

%
\section{Results}
\la{se:numerics}

%
\subsection{Diagrams}

At tree and 1-loop levels the effective potential follows from the Lagrangian
in \eq\nr{V} and from 1-loop contributions in terms of the
sum-integral $J$ defined in appendix~\ref{ss:J}. 
With masses and degeneracies
as listed in table~\ref{table:masses}, we get
\ba
 V^{ }_0 + V^{ }_1 & = & 
 \frac{ \mu_1^2 v^2 }{2} + \frac{ \lambda^{ }_1 v^4 }{4} + 
 \frac{ \delta \mu_1^2 v^2 }{2} + \frac{ \delta \lambda^{ }_1 v^4 }{4}  
 \nn 
 & + & 
 J(\mht) + 3 J(\mGt) + J(\mHt) + J(\mAt) + 2 J(\mHpmt)
 \nn 
 & + & 
 3 \bigl[ (D-3) J(\mW) + J(\mWt)\bigr] + 
 8 \bigl[ (D-3) J(\zerob) + J(\mC)\bigr]
 \nn
 & - & 12 J(\mt^{ } ) - (30\nG -12)J(\zerof)
 \;, \la{V1} \hspace*{8mm}
\ea
where the counterterms are from \eq\nr{ct_h}.
The 2-loop diagrams are given in appendix~B.

%
\subsection{Cancellation of divergences}
\la{ss:cancel2}

The cancellation of divergences through vacuum counterterms offers for
a useful crosscheck of the computation. As illustrated 
in \se\ref{ss:illustration}, the cancellation is non-trivial and  
incomplete in the presence of the thermal resummation introduced in
\eq\nr{resum}. We briefly summarize here the cancellations that 
can be observed. 

First of all, vacuum (i.e.\ temperature independent) divergences are
cancelled by counterterms of a {\em lower} loop order. Parametrically, 
the tree-level potential $V^{ }_0$ is of order $m^4/g^2$. The 1-loop
contribution 
$V^{ }_1$ is of order $m^4$ and contains divergences. 
These are cancelled by tree-level counterterms $\delta Z \sim g^2$, 
which modify $V^{ }_0$ by effects of $\sim V^{ }_0 \, \delta Z \sim m^4$. 
Similarly, $V^{ }_2$ is of order $g^2 m^4$ and contains divergences. 
These are cancelled by contributions of $\sim g^4$ to $\delta Z$
appearing in $V^{ }_0 \sim m^4/g^2$, and by 1-loop effects containing 
the counterterms, likewise of order $V^{ }_1 \delta Z \sim g^2 m^4$.

In contrast, thermal divergences are cancelled by {\em higher-order} effects. 
When the thermal masses of table~\ref{table:masses} are used within
the 1-loop expression, cf.\ \eq\nr{V1}, then the divergent part of 
the function $J$, cf.\ \eq\nr{Jdiv},  leads to temperature-dependent
divergences. Writing $V^{ }_1 = V^{(-1)}_1 / \epsilon + V^{(0)}_1 + \ldots
$ and denoting the thermal part by $V^{(-1)}_{1,\rmii{$T$}}$, we get 
\be
 \frac{V_{1,\rmii{$T$}}^{(-1)}}{\epsilon} = 
 -\frac{ 
 \delta \mphiT^2 \, (\mh^2 + 3 \mG^2) 
 + \delta \mchiT^2 \, (\mH^2 + \mA^2 + 2 \mHpm^2)
 + 3 \mE^2 \mW^2
 }{2(4\pi)^2\epsilon} + (\mbox{$v$-independent})
 \;. \la{V1div}
\ee
Recalling the divergent part of the function $I$ as given in \eq\nr{Idiv}, 
\eq\nr{V1div}
is cancelled by the parts of $V^{ }_2$ given in \eqs\nr{s} and \nr{v} that
contain the thermal counterterms: 
\ba
 (\mbox{s}) + (\mbox{v}) & =  &
 -\fr12 \biggl\{ 
 \delta \mphiT^2 \, \Bigl[ I(\mht) + 3 I(\mGt) \Bigr] + 
 \delta \mchiT^2 \, \Bigl[ I(\mHt) + I(\mAt) + 2 I(\mHpmt) \Bigr]
 \nn & + &  3 \mE^2 I(\mWt)
 \biggr\}
 \;. \la{s+v}
\ea
There is a $v$-independent remainder $\propto \delta \mphiT^4/\epsilon$
left over which is fully cancelled only once the ``3-loop'' diagram (c) 
in \fig\ref{fig:graphs} is included, as discussed in \se\ref{ss:illustration}.

A stringent test is given by the cancellation of non-analytic
divergences, of the type in \eq\nr{non-analytic}. We find that 
divergences proportional to the functions 
$
 I^{(0)}_\rmii{$T$}(\underline{\mW})
$, 
$
 I^{(0)}_\rmii{$T$}(\underline{\mWt})
$, 
$
 I^{(0)}_\rmii{$T$}(\mht)
$, 
$
 I^{(0)}_\rmii{$T$}(\mGt)
$,
$
 I^{(0)}_\rmii{$T$}(\mHt)
$, 
$
 I^{(0)}_\rmii{$T$}(\mAt)
$, 
$
 I^{(0)}_\rmii{$T$}(\mHpmt)
$, 
$
 I^{(0)}_\rmii{$T$}(\mW)
$
and 
$
 I^{(0)}_\rmii{$T$}(\mWt)
$
do cancel, apart from terms proportional to thermal masses, 
which are cancelled by higher-order contributions as discussed
in \se\ref{ss:illustration}.  

In our practical procedure, we let the divergences be 
cancelled by the vacuum counterterms to the extent that this happens. 
The remaining divergences, which are
proportional to thermal masses and thereby formally of higher order, 
are removed by hand. 
Furthermore, because divergences proportional to thermal masses do 
cancel at higher order, we do {\em not} expand the thermal masses
in $\epsilon$, but remove these divergences as a whole. This 
implies, for instance, that the finite part of
the 1-loop contribution in \eq\nr{V1} becomes
\ba
 V^{(0)}_1 & = & 
 J^{(0)}(\mht) + 3 J^{(0)}(\mGt)
 + J^{(0)}(\mHt) + J^{(0)}(\mAt) + 2 J^{(0)}(\mHpmt)
 \nn 
 & + & 
 3 \bigl[ J^{(0)}(\mW) - 2 J^{(-1)}(\mW) + J^{(0)}(\mWt)\bigr] + 
 8 \bigl[ J^{(0)}(\zerob) - 2 J^{(-1)}(\zerob) + J^{(0)}(\mC)\bigr]
 \nn
 & - & 12 J^{(0)}(\mt^{ } ) - (30\nG -12)J^{(0)}(\zerof)
 \;, \la{V1_0} \hspace*{8mm}
\ea
where the functions $J^{(-1)}$ and $J^{(0)}$ are from 
\eqs\nr{Jdiv} and \nr{J0}, respectively. Similarly, 
the 2-loop potential $V^{(2)}$ contains contributions of the types
$I^{(0)}I^{(0)}$, $I^{(-1)}I^{(1)}$, $H^{(0)}$ and, from coefficients 
containing $D = 4 -2\epsilon$, $I^{(-1)}I^{(0)}$ and $H^{(-1)}$.

%
\subsection{Fixing the couplings}

%
\begin{table}[t]

\begin{center}
\begin{tabular}{cccccc}
 \hline \\[-3mm]
 scenario & $\mH/$GeV  & $\mA$/GeV  &  $\mHpm$/GeV & 
 $(\lambda^{ }_3 + \lambda^{ }_4 + \lambda^{ }_5)(\mZ)/2$ &
 $\lambda^{ }_2(\mZ)$ \\ 
 \hline
 BM1 & 66 & 300 & 300 & $\;$1.07$\times 10^{-2}$ & 0.01 \\ 
 BM2 & 200 & 400 & 400 & $\;$1.00$\times 10^{-2}$ & 0.01 \\ 
 BM3 & 5 & 265 & 265 & $-$0.60$\times 10^{-2}\;\,$ & 0.01 \\ 
 \hline 
\end{tabular} 
\end{center}

\vspace*{3mm}

\caption[a]{\small
 The benchmark scenarios from ref.~\cite{pro}. The values of 
 $(\lambda^{ }_3 + \lambda^{ }_4 + \lambda^{ }_5)/2$  and $\lambda^{ }_2$
 refer to the renormalization scale $\bmu = \mZ$. The smallness
 of $(\lambda^{ }_3 + \lambda^{ }_4 + \lambda^{ }_5)/2$ was justified
 with dark matter relic density considerations, and that of 
 $\lambda^{ }_2$ with constraints from dark matter self-interactions. 
 }
\label{table:benchmark}
\end{table}
%

For numerical evaluations we focus on three benchmark points, 
introduced in ref.~\cite{pro}. 
As it turns out, this is sufficient for addressing generic
issues concerning the high-$T$ expansion and the 
convergence of the perturbative expansion. 
The physical parameters associated
with these benchmarks are listed in table~\ref{table:benchmark}. 

%
\begin{table}[t]

\begin{center}
\begin{tabular}{ccccccc}
 \hline \\[-3mm]
 & \multicolumn{2}{c}{BM1}  
 & \multicolumn{2}{c}{BM2}
 & \multicolumn{2}{c}{BM3} \\ 
 & 1-loop & ``2-loop'' 
 & 1-loop & ``2-loop'' 
 & 1-loop & ``2-loop''   \\  
 \hline
 $\mu_1^2(\mZ)/$GeV$^2$ & -6669  & -6568  & -8463  &  -8127  
 & -7392  & -7251 \\ 
 $\mu_2^2(\mZ)/$GeV$^2$ &  842   &  842   & 36620  &  36620  
 & -1243  & -1243 \\ 
 $\lambda_1^{ }(\mZ)$   & 0.0670 & 0.0634 & 0.0671 &  0.0579 
 & 0.1021 & 0.0979 \\ 
 $\lambda_2^{ }(\mZ)$   & 0.010  & 0.010  & 0.010  &  0.010  
 & 0.010  & 0.010 \\ 
 $\lambda_3^{ }(\mZ)$   & 2.757  & 2.757  & 2.618  &  2.618  
 & 2.243  & 2.243 \\ 
 $\lambda_4^{ }(\mZ)$   & -1.368 & -1.368 & -1.299 &  -1.299 
 & -1.127 & -1.127 \\ 
 $\lambda_5^{ }(\mZ)$   & -1.368 & -1.368 & -1.299 &  -1.299 
 & -1.127 & -1.127 \\ 
 $g_2^2(\mZ)$           & 0.425  &  0.425 & 0.425  &  0.425  
 & 0.425  & 0.425 \\ 
 $g_3^2(\mZ)$           & 1.489  &  1.489 & 1.489  &  1.489  
 & 1.489  & 1.489 \\ 
 $h_t^2(\mZ)$           & 0.971  &  0.971 & 0.973  &  0.973  
 & 0.969  & 0.969 \\ 
 \hline 
\end{tabular} 
\end{center}

\vspace*{3mm}

\caption[a]{\small
 Values of $\msbar$ couplings at the scale 
 $\bmu = \mZ$, obtained as explained in appendix~\ref{ss:practice}. 
 Here ``2-loop'' signals that a subset of 2-loop corrections was
 included in $\mu_1^2$ and $\lambda^{ }_1$. 
 For the thermal analysis the values are treated as fixed input, 
 so more digits have been given than are physically accurate. 
 }
\label{table:couplingsmZ}
\end{table}
%

A common feature of all the benchmark points is that the mass
splittings in the inert sector are larger than would ``naturally'' be 
expected from electroweak symmetry breaking, specifically 
$\mA - \mH \gg \mZ$ and $\mHpm - \mH \gg \mZ$. 
This assumption necessitates some of the inert scalar
self-couplings to be large. 
In fact, the couplings are so large that 1-loop corrections 
to physical parameters, such as pole masses, are of order unity. 
The ingredient from vacuum renormalization that is relevant for
our study is the determination of the values of all $\msbar$ parameters 
at some reference scale, chosen as $\bmu = \mZ$
in accordance with ref.~\cite{pro}. The procedure that
we have adopted for estimating these values is to employ a ``self-consistent'' 
prescription in order to resum a subset of higher-order corrections 
and thereby to delimit the magnitude of loop effects;
details are deferred to  
appendix~\ref{ss:practice}.\footnote{%
 To summarize briefly, the Higgs sector parameters 
 $\mu_1^2, \lambda^{ }_1$ are evaluated {\it \`a la} Coleman-Weinberg; 
 the other couplings are evaluated iteratively such that the same values
 appear on both sides of the equation. 
 } 
The resulting couplings are listed 
in table~\ref{table:couplingsmZ}. For the remainder of this 
study, we can forget about vacuum renormalization and simply
use the values in table~\ref{table:couplingsmZ} as input. 
We have checked that variations of the renormalization 
prescription, which lead to 
$\rmO(20\%)$ variations of $\mu_1^2$ and $\lambda^{ }_1$ for BM2, 
nevertheless leave our physics conclusions concerning 
thermal effects qualitatively intact. 

An important first observation from table~\ref{table:couplingsmZ} is 
that the Higgs self-coupling $\lambda^{ }_1$ can be smaller than
in the Standard Model, $\lambda^{ }_1(\mZ) \simeq 0.07 \ll 0.15$. 
A small quartic coupling favours a strong phase 
transition. However, for thermal considerations, the renormalization
scale permitting to avoid large logarithms differs from that used 
for vacuum renormalization. Specifically, thermal fluctuations 
introduce logarithms of the type $\ln(\bmu/(\pi T))$~\cite{generic}. 
Therefore at finite temperature we use 
\be
 \bmu = \alpha \bmu^{ }_T \;, \quad 
 \bmu^{ }_T \equiv \pi T \;, \quad
 \alpha \in (0.5, 2.0)  
 \;. \la{scale}
\ee
The couplings are run between $\bmu = \mZ$ and $\bmu = \alpha \bmu^{ }_T$ 
according to 1-loop renormalization group equations as specified 
in appendix~\ref{ss:RG}. We believe 
that uncertainties from higher-order corrections to the running
are of secondary importance
for the qualitative issues that we are addressing, 
particularly the convergence of the high-$T$ expansion.

%
\subsection{Numerical evaluation}
\la{ss:num}

We have evaluated 
the 1-loop and 2-loop effective potentials both in a closed form
utilizing the high-$T$ expansion, and numerically without
resorting to it. 

As discussed above, the computation is theoretically consistent
only in stable phases: the ``symmetric phase'' at high temperatures
and the ``Higgs phase'' at low temperatures. Outside of these phases the
results are gauge dependent. In addition the results 
contain uncancelled ultraviolet divergences at any finite loop
order as discussed in \ses\ref{ss:illustration} and \ref{ss:cancel2}. 
Once the ultraviolet divergences are removed by hand, an uncancelled
$\bmu$-dependence is left over. Furthermore, 
some masses may become tachyonic; we replace the masses squared by their 
absolute values in such cases. However, these ambiguities 
are numerically benign  
in comparison with the ``physical'' uncertainty associated
with scale dependence, which is formally of higher order but in 
practice substantial, given the large values of some of the mass
parameters and couplings. This uncertainty is estimated through
the scale variation in \eq\nr{scale}. 

\begin{figure}[t]

\hspace*{-0.1cm}
\centerline{%
 \epsfxsize=5.0cm\epsfbox{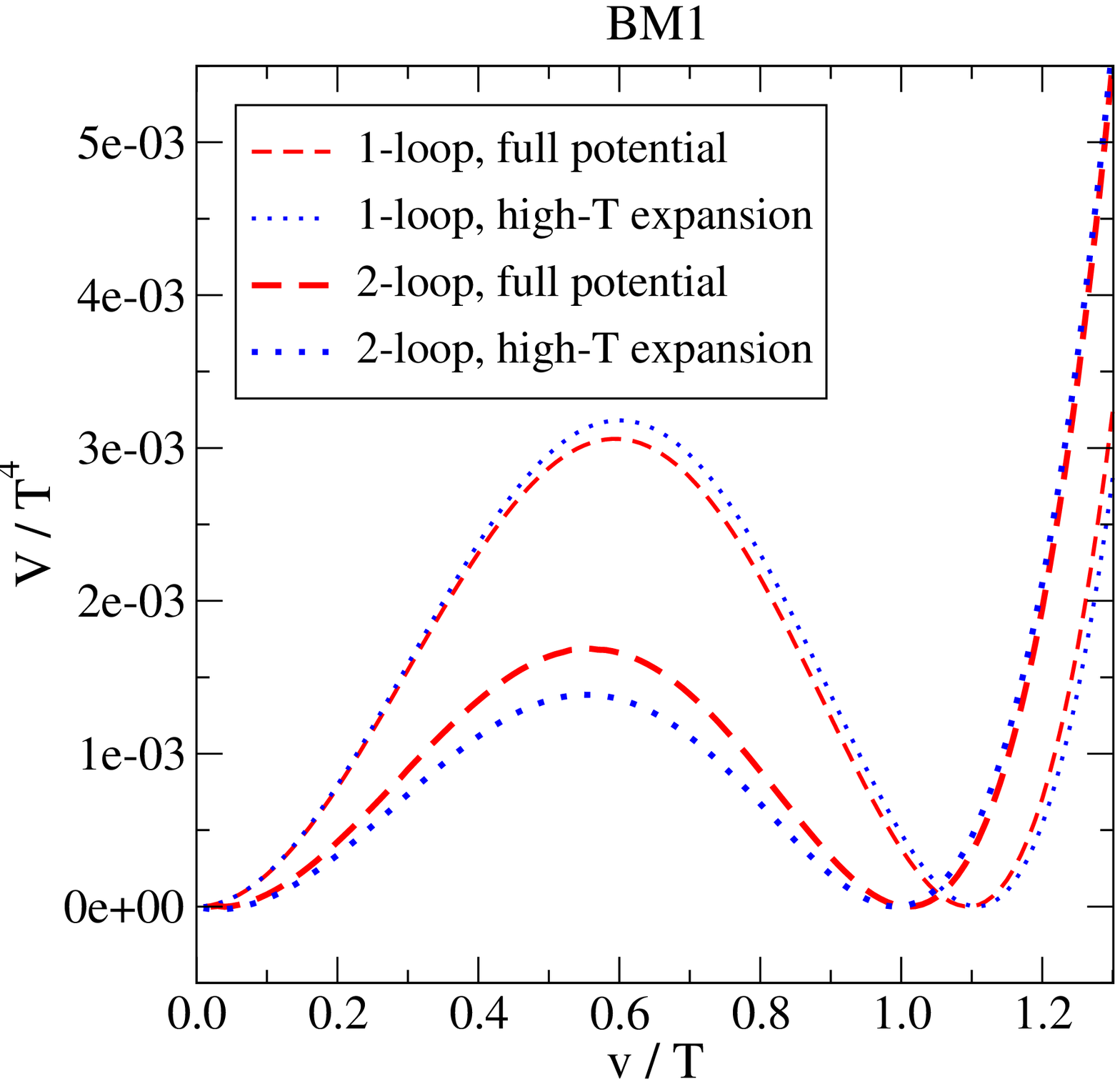}
 \hspace{0.1cm}
 \epsfxsize=5.0cm\epsfbox{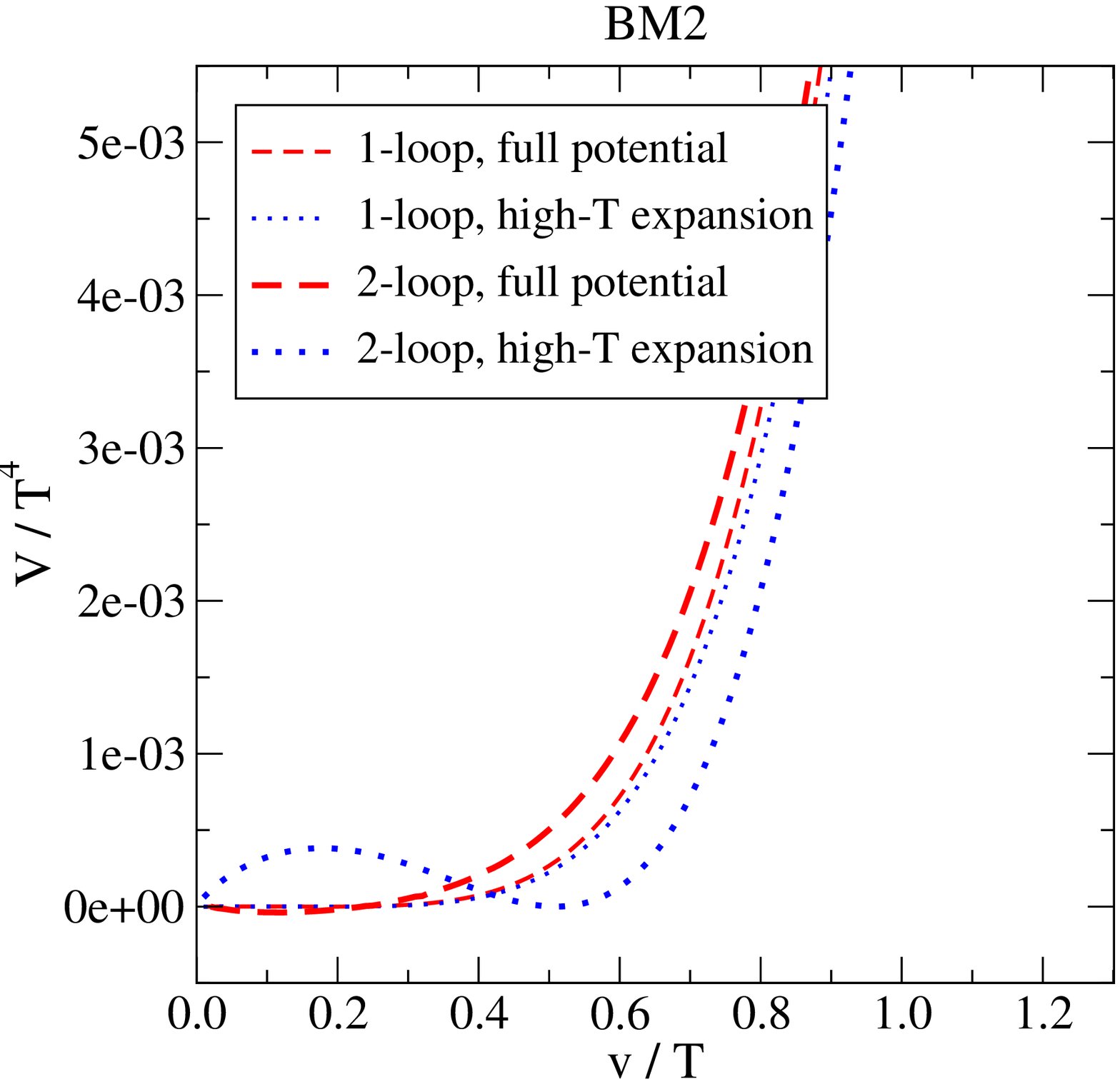}%
 \hspace{0.1cm}
 \epsfxsize=5.0cm\epsfbox{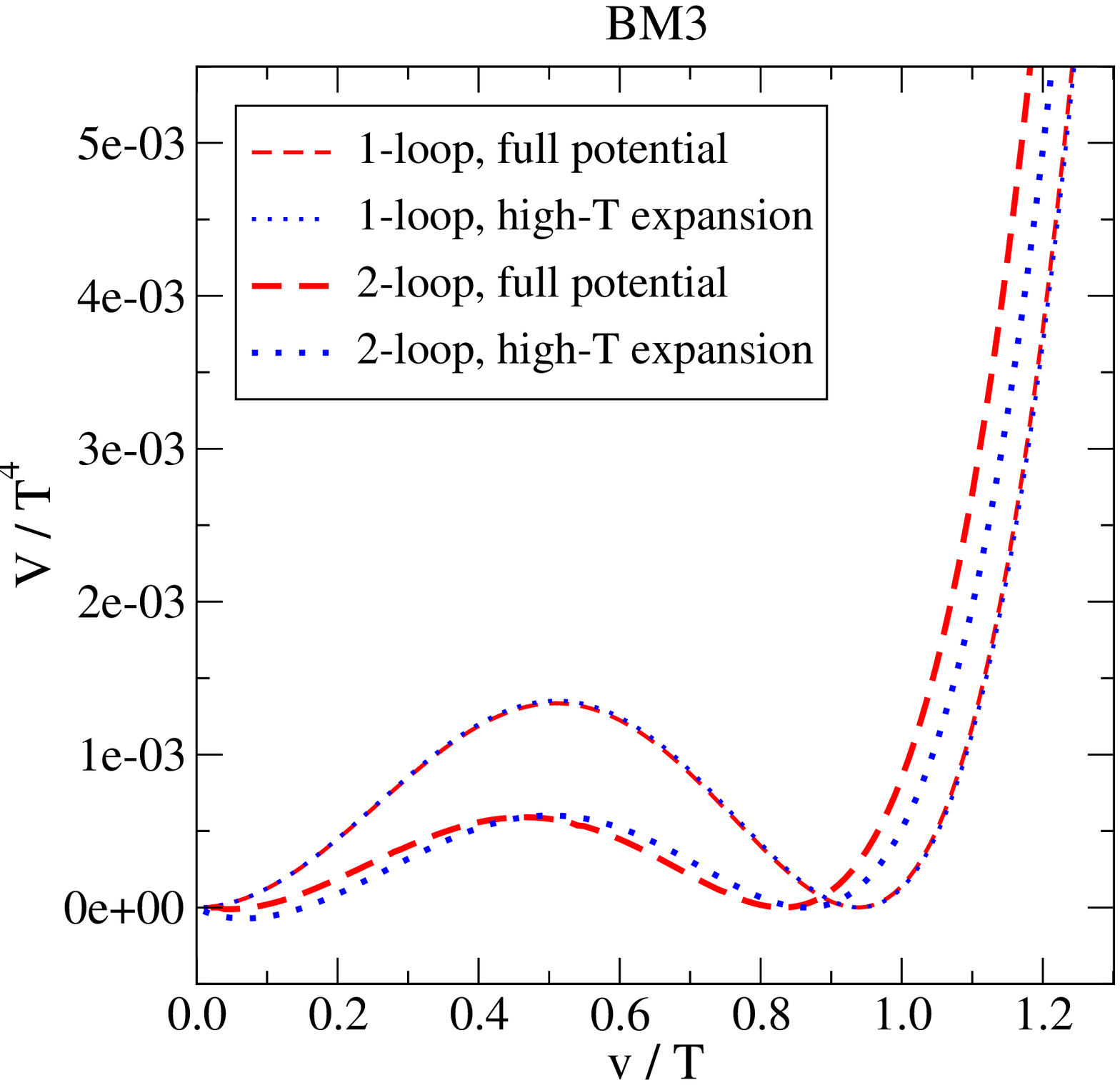}%
}

\caption[a]{\small
 Resummed 1-loop and 2-loop effective potentials, both with 
 and without the high-$T$ expansion, close to the 
 respective critical temperatures, for the various benchmark
 points listed in table~\ref{table:benchmark}. 
 We stress that, as elaborated upon in \ses\ref{se:renorm}
 and \ref{ss:cancel2}, the effective potential is gauge dependent and
 in the presence of resummation contains uncancelled
 divergences at any finite loop order, with ``loop order'' defined in the 
 sense of \fig\ref{fig:graphs}; hence the plots are meant for  
 illustration only. The gauge-independent results that can be derived
 from the effective potential 
 are given in table~\ref{table:benchmark2}.
}

\la{fig:potential}
\end{figure}

We assume in this paper that the Z(2) symmetry 
related to the Inert Doublet $\chi$ is unbroken. In order to verify
the validity of this assumption, the effective mass squared of the $\chi$
field, $\mu_2^2 + \delta \mchiT^2$ in the notation 
of table~\ref{table:masses}, is evaluated at the critical temperature.
We find $ \mu_2^2 + \delta \mchiT^2 \,\gsim\, (0.6 T)^2$ in all 
cases, justifying the assumption {\it a posteriori}, given that
$0.6 T \sim gT$ is parametrically a ``heavy'' scale, similar to $\mE$. 

In \fig\ref{fig:potential}, the 1-loop and 2-loop effective potentials
are plotted at the corresponding critical temperatures, 
both with and without a high-$T$ expansion. We note that 
the 2-loop corrections are substantial but in general they 
do not modify the 1-loop predictions qualitatively. 
For BM1 and BM3 they weaken the transition moderately, and for 
BM2 they remove the transition altogether for $\bmu \gsim \bmu^{ }_T$ 
(in contrast they appear to
strengthen the transition for BM2 within the 
high-$T$ expansion, 
however those results are unreliable because of the large
value of $\mu_2^2$). 
With the exception of BM2, 
the high-$T$ expansion is seen to work very well. 
However, the value of the critical temperature
does change substantially through the high-$T$ 
expansion; this is easily understood and
is elaborated upon in \se\ref{se:concl}. 

For BM2, in which case we find a very weak transition, the problems
mentioned at the end of \se\ref{se:resum}, associated on one hand 
with the breakdown of the high-$T$ expansion for heavy inert modes
(cf.\ table~\ref{table:couplingsmZ}), 
and on the other 
with infrared sensitive terms not captured by our simple thermal 
resummation, 
become visible as ``linear terms'' at small $v/T$. 
We do not elaborate on these 
any further, given that 
the transition is too weak to be of physical interest. 

Physical (gauge independent) results for the 
quantities $\Tc,L$ and $v^{ }_\rmi{phys}$, defined in \se\ref{se:model},
are collected in table~\ref{table:benchmark2}. The errors originate
from a variation of $\bmu$ in the range $(0.5 ... 2.0)\, \pi T$,
cf.\ \eq\nr{scale}. These results confirm the heuristic 
impressions visible in \fig\ref{fig:potential}. At least
for BM1, the transition could be marginally
strong enough to support electroweak baryogenesis. 

%
\begin{table}[t]

\begin{center}
\begin{tabular}{ccccccccc}
 & & & 
 \multicolumn{4}{c}{full effective potential}  
 & &    
 \\[1mm] \hline 
 & \multicolumn{2}{c}{$\Tc/\mbox{GeV}$}  
 & \multicolumn{2}{c}{$L/\Tc^4$}
 & \multicolumn{2}{c}{$v^{ }_\rmi{phys}/\Tc$}  
 & \multicolumn{2}{c}{$v^{ }_\rmi{min}/\Tc$} \\ 
 & 1-loop & 2-loop 
 & 1-loop & 2-loop 
 & 1-loop & 2-loop   
 & 1-loop & 2-loop   \\  
 \hline
 BM1 & 139(14) & 155(21) & 0.44(1) & 0.34(1)
     & 1.14(12) & 0.98(4) & 1.15(12) & 0.98(3)  \\ 
 BM2 & 159(13) & 181(22) & 0.07(7) & 0.03(3)
     & 0.39(28) & 0.16(16) & 0.39(28) & 0.17(17) \\ 
 BM3 & 138(8) & 167(19) & 0.35(3) & 0.20(1)
     & 0.96(10) & 0.84(6) & 0.98(10)  & 0.81(2) \\ 
 \hline 
\end{tabular} 
\end{center}

\vspace*{3mm}

\begin{center}
\begin{tabular}{ccccccccc}
 & & & 
 \multicolumn{4}{c}{high-$T$ expansion}  
 & &    
 \\[1mm] \hline 
 & \multicolumn{2}{c}{$\Tc/\mbox{GeV}$}  
 & \multicolumn{2}{c}{$L/\Tc^4$}
 & \multicolumn{2}{c}{$v^{ }_\rmi{phys}/\Tc$}  
 & \multicolumn{2}{c}{$v^{ }_\rmi{min}/\Tc$} \\ 
 & 1-loop & 2-loop 
 & 1-loop & 2-loop 
 & 1-loop & 2-loop   
 & 1-loop & 2-loop   \\  
 \hline
 BM1 & 140(14)  & 124(8) & 0.45(1)  & 0.49(22)  
     & 1.15(13) & 1.04(31) & 1.16(13) & 1.05(32) \\  
 BM2 & 159(14) & 140(9) & 0.08(8)  & 0.16(8)
     & 0.42(30) & 0.60(19) & 0.42(30) & 0.60(19) \\  
 BM3 & 138(8) & 125(3) & 0.35(3)  & 0.37(16)
     & 0.97(10) & 0.89(23) & 0.98(10)  & 0.91(23) \\  
 \hline 
\end{tabular} 
\end{center}

\vspace*{3mm}

\caption[a]{\small
 Results for the physical quantities defined in \se\ref{se:model},
 as well as for the gauge dependent $v^{ }_\rmi{min}$ evaluated 
 in Feynman $R^{ }_\xi$ gauge, 
 for the benchmark scenarios listed in table~\ref{table:benchmark}.
 The uncertainties were obtained through the scale variation
 in \eq\nr{scale}. 
 We note that the scale uncertainties, which reflect the size of 
 higher-order corrections from large scalar couplings,
 completely dominate over ambiguities related to gauge dependence, 
 whose size is indicated by the difference between $v^{ }_\rmi{phys}$ and
 $v^{ }_\rmi{min}$.  
 }
\label{table:benchmark2}
\end{table}
%

%
\section{Conclusions}
\la{se:concl}

In this paper we have developed the general technology for evaluating
the full 2-loop thermal effective potential for the Higgs field, 
without resorting to a high-temperature expansion. The technology
has been applied to  
the Inert Doublet Model (IDM), incorporating  thermal resummation
in a particularly simple way. 

Even though we did not dwell on this in the text, 
the largest technical effort of our work 
went into the derivation of the formulae 
given in appendices A and B, and into their numerical evaluation, 
which poses a challenge of its own
(some remarks can be found in a paragraph below \eq\nr{Hbar_full}). 
The results of 
appendix~A are model-independent, and can be applied  
to any extension of the Standard Model.  

A word of caution is in order on 
the simple resummation that we adopted (\se\ref{se:resum}). 
As explained at the end of \se\ref{se:resum} 
and in \se\ref{ss:num}, the resummation is not suitable 
for a precise treatment of infrared effects;
at the same time, as explained in \ses\ref{ss:illustration} 
and \ref{ss:cancel2}, 
it is also problematic in the ultraviolet, introducing 
spurious divergences which are only cancelled
by higher-order diagrams. We therefore do not endorse its
use for {\em practical computations} aiming at physical 
precision; for us it was just a simple tool permitting
to compare two different computations
(a full 2-loop analysis and its high-$T$ expansion). 
As explained below, our final conclusion concerning the 
high-$T$ expansion suggests the availability of {\em other tools}
for addressing physical observables with good precision. 

Applying the formalism to the IDM as an example, 
our main finding is that the high-$T$ expansion works well 
for describing the strength of the phase transition, despite the 
fact that some degrees of freedom become heavy in the Higgs phase
(cf.\ fig.~\ref{fig:potential} and table~\ref{table:benchmark2}).
This is a welcome observation, given that it opens the avenue
for dimensionally reduced lattice investigations, necessary 
for cases in which a good precision is needed and/or 
properties associated with inhomogeneous configurations are of interest. 
This concerns for instance the surface tension~\cite{2hdm,mH126},  
the bubble nucleation rate~\cite{bubble}, 
and the sphaleron rate~\cite{sphaleron1}. 

Based on the benchmark points considered, as well as on a 
parameter scan at 1-loop level, we find that in general the IDM transition 
is at most moderately strong, as long as the $\chi$-field does not 
become so light that it would experience a transition of its own. 
In other models, possessing a stronger transition, the high-$T$
expansion could fail. At the same time, 
we would expect smaller 2-loop corrections in those cases, given 
that the infrared sensitive expansion parameter is 
$\sim g^2 T / (\pi\mW) \sim 2 g T/(\pi v) $. 

There is one observable for which the high-$T$
expansion does {\em not} work well: the critical
temperature $\Tc$ 
(cf.\ the 2-loop results in table~\ref{table:benchmark2}). This should
not come as a surprise: some of the inert scalars {\em are} heavy and/or
strongly coupled, 
and should not be treated with the high-$T$ expansion nor
with the leading-order resummation of \se\ref{se:resum}. 
Even though they have little effect on the phase transition, they
``renormalize'' the effective Higgs mass parameter by a large amount. 
Within a dimensionally reduced investigation~\cite{dr1,dr2}, 
these effects can be 
incorporated without a high-$T$ expansion~\cite{old} and 
including higher orders in large couplings~\cite{generic}, 
so a good precision can be expected also for $\Tc$. 
We believe that a detour through the dimensionally reduced 
description, with 
effects of $\rmO(\lambda^{ }_3 \mu_2^2)$  and $\rmO(\lambda_3^2 T^2)$ 
included in thermal masses,  
should be chosen even in purely perturbative studies, 
if numerical precision at or below the 10\% level is needed.  

Turning finally to cosmology, our study supports previous
suggestions according to which the IDM can incorporate
a phase transition marginally strong enough for baryogenesis, at least for the 
benchmark point BM1 (cf.\ table~\ref{table:benchmark2}). The 2-loop
corrections weaken the transition somewhat, but in many cases they also 
reduce the scale uncertainties of $L,v^{ }_\rmi{phys}$
and $v^{ }_\rmi{min}$ 
(cf.\ the panel ``full effective potential'' 
in table~\ref{table:benchmark2}).  
However the problem of obtaining sufficient CP violation 
is not alleviated by the IDM Higgs sector which contains no new
physical phases. As an example of a work-around, 
it has been suggested
that CP violation in the interactions responsible for 
neutrino masses could play a role for baryogenesis 
(cf.\ e.g.\ ref.~\cite{mrm}).

%
\section*{Acknowledgements}

This work was supported by the Swiss National Science Foundation
(SNF) under grant 200020-168988, 
and by the Munich Institute for Astro- and Particle
Physics (MIAPP) of the DFG cluster of excellence 
``Origin and Structure of the Universe''.

%
\appendix
\renewcommand{\thesection}{\Alph{section}}
\renewcommand{\thesubsection}{\Alph{section}.\arabic{subsection}}
\renewcommand{\theequation}{\Alph{section}.\arabic{equation}}

%
\section{Thermal master sum-integrals}
\la{se:masters}

We list here the expressions for the sum-integrals that are
needed for evaluating the 2-loop effective potential. 
The results are divergent for $\epsilon\to 0$. For the 
1-loop structures the divergences are proportional to $1/\epsilon$; 
the 2-loop result contains squares of 1-loop structures, so that the 
1-loop structures need to be evaluated up to terms of $\rmO(\epsilon)$. 
For genuine 2-loop structures, the terms that need to be tracked are of 
$\rmO(1/\epsilon^2)$, $\rmO(1/\epsilon)$, and $\rmO(1)$.

The master sum-integrals contain both vacuum
(i.e.\ temperature independent) and thermal parts. 
In the following, the thermal corrections are given both 
in an exact form suitable for numerical evaluation, and 
analytically in a high-$T$ expansion. In the latter case, 
the leading contribution is of $\rmO(T^4)$, and we list terms up 
to $\rmO(m^4, g^2 m^2 T^2)$, where $m^2$ denotes a generic mass
squared and $g^2$ a generic coupling constant. This is consistent with 
a power counting $m^2 \lsim g^2 T^2$ which can be used for 
justifying the high-$T$ expansion. Some of the master
structures are always multiplied by $\sim g^2 m^2$, in which case
high-$T$ expansions are given up to $\rmO(T^2)$. 

Thermal corrections depend on whether the particle in question
is a boson or a fermion. In order to compactify the expressions, we employ
an implicit notation in which the statistics is 
identified through the mass carried by the particle. The distribution 
function is denoted generically by $n$, and if the argument is ``bosonic'', 
it is to be interpreted as the Bose distribution,
$n(\E) \to \nB{}(\E)
  \equiv 1 / (e^{\E/T} - 1)$. In contrast, with 
a ``fermionic'' argument, minus the Fermi distribution is 
to be understood, 
$n(\E) \to - 
 \nF{}(\E) \equiv - 1 / (e^{\E/T} + 1)$.
A vanishing bosonic mass is denoted
by $\zerob$ and a fermionic one by $\zerof$.

%
\subsection{Function $J(m)$}
\la{ss:J}

The master sum-integral appearing in the 1-loop result is denoted by
\be
 J(m^{ }) \; \equiv \; 
 \fr12 \Tint{P} \ln \bigl( {P^2 + m^2} \bigr)
 \; = \; 
 \frac{ m^2 A(m^{ }) }{D} 
  - \frac{1}{D-1} \int_{\vec{p}} \frac{p^2 n(\E^{ })}{\E^{ }}
 \;, \la{J}
\ee
where $P = (\omega^{ }_n,\vec{p})$; $\omega^{ }_n$ denote
Matsubara frequencies; $p\equiv |\vec{p}|$; 
the vacuum function $A$ is given in \eq\nr{A}; 
$D = 4 -2\epsilon$;
$\E^{ }_i \equiv \sqrt{p^2 + m_i^2}$; and we made use of 
partial integrations and properties
of dimensional regularization.
We write
\be
 J(m) = \frac{ 1 }{\epsilon} \, J^{(-1)}_{ }(m) 
  + J^{(0)}_{ }(m) 
      + \rmO(\epsilon)
 \;. \la{Jeps}
\ee
Suppressing an overall $\mu^{-2\epsilon}$, where $\mu$ is a scale 
parameter related to dimensional regularization, and denoting
\be
 \ln\bmu^2 \equiv \ln\mu^2 + \ln(4\pi) - \gammaE
 \;, \la{msbar}
\ee
the expressions for the functions in \eq\nr{Jeps} read
\ba
 J^{(-1)}_{ }(m) & = & 
 -\frac{m^4}{4(4\pi)^2} 
 \;, \la{Jdiv} \\ 
 J^{(0)}_{ }(m) & = & 
 -\frac{m^4}{4(4\pi)^2}
 \biggl( \ln\frac{\bmu^2}{m^2} + \fr32 \biggr) 
 - \frac{ I^{(0)}_\rmii{$T$}(\underline{m}) }{3}
 \;, \la{J0}  
 \hspace*{4mm} 
\ea
where $I^{(0)}_\rmii{$T$}(\underline{m})$ is given in \eq\nr{ITmbar}. 
In the high-$T$ limit, the expressions depend on whether bosonic
or fermionic particles are considered. Expanding up to 
$\rmO(m^4)$, the bosonic contributions read
\ba
 J^{(0)}_{ }(\mb) & = & 
 -\frac{\pi^2 T^4}{90}
 + \frac{\mb^2 T^2}{24}
 - \frac{\mb^3 T}{12\pi}
 - \frac{\mb^4 }{2(4\pi)^2}
   \ln\biggl( \frac{\bmu e^{\gammaE}}{4\pi T} \biggr)
 + \rmO\Bigl( \frac{m^6_b}{\pi^4 T^2} \Bigr) 
 \;,  
\ea 
whereas the fermionic expression is  
\ba
 J^{(0)}_{ }(\mf) & = & 
 \fr78 \frac{\pi^2 T^4}{90}
 - \frac{\mf^2 T^2}{48}
 - \frac{\mf^4 }{2(4\pi)^2}
   \ln\biggl( \frac{\bmu e^{\gammaE}}{\pi T} \biggr)
 + \rmO\Bigl( \frac{\mf^6}{\pi^4 T^2} \Bigr) 
 \;.  
\ea

%
\subsection{Function $I(m)$}
\la{ss:I}

The basic 1-loop structure appearing within the 2-loop result is denoted by
\be
 I(m^{ }) \; \equiv \; 
 \Tint{P} \frac{1}{P^2 + m^2}
 \; = \; 
 A(m^{ }) + \int_{\vec{p}} \frac{n(\E^{ })}{\E^{ }}
 \;, \la{I}
\ee
where the vacuum part $A$ is from \eq\nr{A}.
We write
\be
 I(m) = \frac{ 1 }{\epsilon} \, I^{(-1)}_{ }(m) 
  + I^{(0)}_{ }(m) + \epsilon\, I^{(1)}_{ }(m) 
      + \rmO(\epsilon^2)
 \;, \la{Ieps}
\ee
and subsequently separate each contribution into a vacuum and thermal part, 
\be
 I^{(n)}_{ }(m) = 
 I^{(n)}_{0}(m) + 
 I^{(n)}_\rmii{$T$}(m) \;. 
\ee
Suppressing an overall $\mu^{-2\epsilon}$, 
the expressions for the functions in \eq\nr{Ieps} read
\ba
 I^{(-1)}_{ }(m) & = & 
 -\frac{m^2}{(4\pi)^2} 
 \;, \la{Idiv} \\[2mm] 
 I^{(0)}_{ }(m) & = & 
 -\frac{m^2}{(4\pi)^2}
 \biggl( \ln\frac{\bmu^2}{m^2} + 1 \biggr) 
 + I^{(0)}_\rmii{$T$}(m) \;, \la{I0} \\[2mm] 
 I^{(0)}_\rmii{$T$}(m) & = &  
 \int_0^\infty \frac{{\rm d} p \, p^2 \, n(\E)}{2\pi^2 \E} 
 \;, \\[2mm] 
 I^{(1)}_{ }(m) & = & 
 -\frac{m^2}{(4\pi)^2}
 \biggl( \fr12 \ln^2 \frac{\bmu^2}{m^2} + \ln \frac{\bmu^2}{m^2} 
 + 1 + \frac{\pi^2}{12} \biggr) 
 +  I^{(1)}_\rmii{$T$}(m) \;, \\[2mm] 
 I^{(1)}_\rmii{$T$}(m) & = & 
 \int_0^\infty \frac{{\rm d} p \, p^2 \, (\ln\frac{\bmu^2}{4 p^2} + 2)
  n(\E)}{2\pi^2 \E}
 \;.  
 \hspace*{4mm} 
\ea
These functions are related through
$
 I^{(n)}_{ }(m) = \bmu^2 \partial I^{(n+1)}_{ }(m) / \partial \bmu^2
$.
In the high-$T$ limit, the bosonic contributions read
\ba
 I^{(0)}_{ }(\mb) & = & 
 \frac{T^2}{12} - \frac{\mb T}{4\pi}
 - \frac{2 \mb^2}{(4\pi)^2} 
 \ln\biggl( \frac{\bmu e^{\gammaE}}{4\pi T}\biggr) 
 + \rmO\Bigl( \frac{m^4_b}{\pi^4 T^2} \Bigr) 
 \;, \\ 
 I^{(1)}_{ }(\mb) & = & 
 \frac{T^2}{6} \, \biggl[
  \ln\biggl( \frac{\bmu e^{\gammaE}}{2 T}\biggr) - \frac{\zeta'(2)}{\zeta(2)} 
  \biggr]
 - \frac{\mb T}{2\pi}
   \biggl[ \ln\Bigl( \frac{\bmu}{2 \mb} \Bigr) + 1 \biggr] 
 \nn 
 & - &  \frac{2 \mb^2}{(4 \pi)^2}
 \biggl[
 \ln^2\biggl( \frac{\bmu e^{\gammaE}}{4\pi T}\biggr)
 - \gammaE^2 
 - 2 \gamma^{ }_1 + \frac{\pi^2}{8}
 \biggr] 
 \; + \; \rmO\Bigl( \frac{m^4_b}{\pi^4 T^2} \Bigr) 
 \;,  \la{I1b}
\ea 
where the Stieltjes constant $\gamma_1$ is  
defined through $\zeta(s) = 1/(s-1) + 
\sum_{n = 0}^\infty \gamma_n (-1)^n (s - 1)^n/n!$.
The corresponding fermionic expressions are 
\ba
 I^{(0)}_{ }(\mf) & = & 
 - \frac{T^2}{24} 
 - \frac{2 \mf^2}{(4 \pi)^2} 
 \ln\biggl( \frac{\bmu e^{\gammaE}}{\pi T}\biggr) 
 + \rmO\Bigl( \frac{\mf^4}{\pi^4 T^2} \Bigr) 
 \;, \\ 
 I^{(1)}_{ }(\mf) & = & 
 - \frac{T^2}{12} \, \biggl[
  \ln\biggl( \frac{\bmu e^{\gammaE}}{4 T}\biggr) - \frac{\zeta'(2)}{\zeta(2)} 
  \biggr]
 \nn 
 & - &  \frac{2 \mf^2}{(4 \pi)^2}
 \biggl[
 \ln^2\biggl( \frac{\bmu e^{\gammaE}}{\pi T}\biggr)
 - \gammaE^2
 - 2 \ln^2\bigl(2\bigr)
 - 2 \gamma^{ }_1 + \frac{\pi^2}{8}
 \biggr] 
 \; + \; \rmO\Bigl( \frac{\mf^4}{\pi^4 T^2} \Bigr) 
 \;.  \hspace*{5mm} \la{I1f}
\ea 

%
\subsection{Function $I(\underline{m})$}
\la{ss:Ibar}

The thermal 2-loop effective potential contains appearances of a 
``Lorentz-violating'' 1-loop structure denoted by
\be
 I(\underline{m^{ }}) \; \equiv \; 
 \Tint{P} \frac{p^2}{P^2 + m^2}
 \; = \; 
 - \frac{D-1}{D} m^2A(m^{ })
  + \int_{\vec{p}} \frac{p^2 n(\E^{ })}{\E^{ }}
 \;, \la{Ibar}
\ee
where the vacuum part $A$ is from \eq\nr{A}.
We write
\be
 I(\underline{m}) = \frac{ 1 }{\epsilon} \, I^{(-1)}_{ }(\underline{m}) 
  + I^{(0)}_{ }(\underline{m}) + \epsilon\, I^{(1)}_{ }(\underline{m}) 
      + \rmO(\epsilon^2) \;,
 \la{Ibar_eps}
\ee
and 
$
 I^{(n)}_{ }(\underline{m}) = 
 I^{(n)}_{0}(\underline{m}) + 
 I^{(n)}_\rmii{$T$}(\underline{m})  
$.
Suppressing an overall $\mu^{-2\epsilon}$, 
the expressions for the structures in \eq\nr{Ibar_eps} read
\ba
 I^{(-1)}_{ }(\underline{m}) & = & 
 \frac{3 m^4}{4(4\pi)^2} 
 \;, \\[2mm] 
 I^{(0)}_{ }(\underline{m}) & = & 
 \frac{m^4}{(4\pi)^2}
 \biggl(\fr34 \ln\frac{\bmu^2}{m^2} + \fr58 \biggr) 
 + I^{(0)}_\rmii{$T$}(\underline{m}) 
 \;, \\[2mm] 
 I^{(0)}_\rmii{$T$}(\underline{m}) & = & 
 \int_0^\infty \frac{{\rm d} p \, p^4 \, n(\E)}{2\pi^2 \E} 
 \;,  \la{ITmbar}  \\[2mm] 
 I^{(1)}_{ }(\underline{m}) & = & 
 \frac{m^4}{(4\pi)^2}
 \biggl( \fr38 \ln^2 \frac{\bmu^2}{m^2} + \fr58 \ln \frac{\bmu^2}{m^2} 
 + \frac{9}{16} + \frac{\pi^2}{16} \biggr) 
 + I^{(1)}_\rmii{$T$}(\underline{m}) 
 \;, \\[2mm] 
 I^{(1)}_\rmii{$T$}(\underline{m}) & = & 
 \int_0^\infty \frac{{\rm d} p \, p^4 \, (\ln\frac{\bmu^2}{4 p^2} + 2)
  n(\E)}{2\pi^2 \E}
 \;.  
\ea
These functions are related through
$
 I^{(n)}_{ }(\underline{m}) = \bmu^2 \partial 
 I^{(n+1)}_{ }(\underline{m}) / \partial \bmu^2
$.
In the high-$T$ limit, the bosonic contributions read
\ba
 I^{(0)}_{ }(\underline{\mb}) & = & 
 \frac{\pi^2 T^4}{30} 
 - \frac{\mb^2 T^2}{8} 
 + \rmO\Bigl( \frac{m^3_b T}{\pi} \Bigr) 
 \;, \\ 
 I^{(1)}_{ }(\underline{\mb}) & = & 
 \frac{\pi^2 T^4}{15} \, \biggl[
  \ln\biggl( \frac{\bmu e^{\gammaE}}{2 T}\biggr) - \frac{\zeta'(4)}{\zeta(4)} 
  - \fr56
  \biggr]
 \nn 
 &  - &
 \frac{\mb^2 T^2}{4} \, \biggl[
  \ln\biggl( \frac{\bmu e^{\gammaE}}{2 T}\biggr) - \frac{\zeta'(2)}{\zeta(2)} 
  - \fr13
  \biggr]
 + \rmO\Bigl( \frac{m^3_b T}{\pi} \Bigr) 
 \;,  \la{I1bbar}
\ea 
whereas the corresponding fermionic expressions are 
\ba
 I^{(0)}_{ }(\underline{\mf}) & = & 
 - \frac{7 \pi^2 T^4}{240} 
 + \frac{\mf^2 T^2}{16} 
 + \rmO\Bigl( \frac{\mf^4}{\pi^4 T^2} \Bigr) 
 \;, \\ 
 I^{(1)}_{ }(\underline{\mf}) & = & 
 - \frac{7 \pi^2 T^4}{120} 
 \, \biggl[
  \ln\biggl( \frac{\bmu e^{\gammaE}}{2 T}\biggr) - \frac{\zeta'(4)}{\zeta(4)} 
  - \frac{\ln2}{7} - \fr56
  \biggr]
 \nn 
 &  + &
 \frac{\mf^2 T^2}{8} \, \biggl[
  \ln\biggl( \frac{\bmu e^{\gammaE}}{4 T}\biggr) - \frac{\zeta'(2)}{\zeta(2)} 
  - \fr13
  \biggr]
 + \rmO\Bigl( \frac{\mf^4}{\pi^2} \Bigr) 
 \;.  \la{I1fbar}
\ea 

%
\subsection{Function $H(m^{ }_1, m^{ }_2, m^{ }_3)$}
\la{ss:H}

Next we consider the 2-loop ``sunset'' sum-integral, 
\be
 H(m^{ }_1, m^{ }_2, m^{ }_3)
 \; \equiv \; 
 \Tint{P,Q} \frac{1}{(P^2 + m_1^2)(Q^2 + m_2^2) [(P+Q)^2 + m_3^2]}
 \;. \la{H}
\ee
The Matsubara sums can be carried out explicitly, and for the vacuum
part the spatial integrations are also doable~\cite{fjj,dt,cclr}.  
Writing the result as 
\be
 H(\{m^{ }_i\})
 \; = \; 
 \frac{1}{\epsilon^2} \, H^{(-2)}_{ }(\{m^{ }_i\})
 +   
 \frac{1}{\epsilon} \, H^{(-1)}_{ }(\{m^{ }_i\})
 +   
  H^{(0)}_{ }(\{m^{ }_i\}) 
 +   
 \rmO(\epsilon)
 \;, \la{H_expand}
\ee
and omitting an overall $\mu^{-4\epsilon}$, the results read
\ba
 H^{(-2)}_{ }(\{m^{ }_i\}) & = & 
 - \frac{m_1^2 + m_2^2 + m_3^2}{2(4\pi)^4}
 \;, \la{Hdiv2} \\ 
 H^{(-1)}_{ }(\{m^{ }_i\}) & = & 
 - \frac{1}{(4\pi)^4} \sum_{i=1}^3 m_i^2 
 \, \Bigl( \ln\frac{\bmu^2}{m_i^2} + \fr32  \Bigr)
 \; + \;  
 \sum_{i=1}^3 \frac{I^{(0)}_\rmii{$T$}(m^{ }_i)}{(4\pi)^2}
 \;, \la{Hdiv1} \\ 
 H^{(0)}_{ }(\{m^{ }_i\}) & = & 
 \frac{1}{(4\pi)^4}
 \biggl\{ 
  -\fr12 \sum_{i=1}^3 m_i^2 \ln^2\Bigl( \frac{\bmu^2}{m_i^2} \Bigr)
  - 3 \sum_{i=1}^3 m_i^2 \ln\Bigl( \frac{\bmu^2}{m_i^2} \Bigr)
  - \Bigl( \fr72 + \frac{\pi^2}{12} \Bigr) \sum_{i=1}^3 m_i^2
 \nn 
 & - & 
 \fr12 
 \biggl[ 
  (m_1^2 + m_2^2 - m_3^2)\,
    \ln\Bigl( \frac{\bmu^2}{m_1^2} \Bigr)
    \ln\Bigl( \frac{\bmu^2}{m_2^2} \Bigr)
 + (m_1^2 + m_3^2 - m_2^2 )\,
    \ln\Bigl( \frac{\bmu^2}{m_1^2} \Bigr)
    \ln\Bigl( \frac{\bmu^2}{m_3^2} \Bigr)
 \nn 
 & + & 
   (m_2^2 + m_3^2 - m_1^2 )\,
    \ln\Bigl( \frac{\bmu^2}{m_2^2} \Bigr)
    \ln\Bigl( \frac{\bmu^2}{m_3^2} \Bigr)
 \biggr]
 +  R(m_1^2,m_2^2,m_3^2)\, L(m_1^2,m_2^2,m_3^2) 
 \biggr\} 
 \nn 
 & + & 
  I^{(0)}_\rmii{$T$}(m^{ }_1)\, 
 \re B^{(0)}_{}(-i m^{ }_1;m^{ }_2, m^{ }_3)
 \; + \; 
  I^{(0)}_\rmii{$T$}(m^{ }_2)\, 
 \re B^{(0)}_{}(-i m^{ }_2;m^{ }_3, m^{ }_1)
 \nn 
 & + & 
  I^{(0)}_\rmii{$T$}(m^{ }_3)\, 
 \re B^{(0)}_{}(-i m^{ }_3;m^{ }_1, m^{ }_2)
 \; + \;
 \sum_{i=1}^3 \frac{I^{(1)}_\rmii{$T$}(m^{ }_i)}{(4\pi)^2}
 \nn 
 & + & 
 \int_{0}^\infty \!  
 \frac{{\rm d}p\, {\rm d} q\, p q\, n(\E^{p}_{1}) n(\E^{q}_{2})}
 {32\pi^4 \E^{p}_{1} \E^{q}_{2}} 
 \ln \biggl| 
  \frac{(m_3^2 - m_1^2 - m_2^2 + 2 p q)^2 - 4 (\E^{p}_{1})^2 (\E^{q}_{2})^2 }
       {(m_3^2 - m_1^2 - m_2^2 - 2 p q)^2 - 4 (\E^{p}_{1})^2 (\E^{q}_{2})^2 }
 \biggr|
 \nn[2mm] 
 & + & 
 \int_{0}^\infty \!  
 \frac{{\rm d}p\, {\rm d} q\, p q\,  n(\E^{p}_{2}) n(\E^{q}_{3})}
 {32\pi^4  \E^{p}_{2} \E^{q}_{3}} 
 \ln \biggl| 
  \frac{(m_1^2 - m_2^2 - m_3^2 + 2 p q)^2 - 4 (\E^{p}_{2})^2 (\E^{q}_{3})^2 }
       {(m_1^2 - m_2^2 - m_3^2 - 2 p q)^2 - 4 (\E^{p}_{2})^2 (\E^{q}_{3})^2 }
 \biggr|
 \nn[2mm] 
 & + & 
 \int_{0}^\infty \!  
 \frac{{\rm d}p\, {\rm d} q\, p q\,  n(\E^{p}_{3}) n(\E^{q}_{1})}
 {32\pi^4  \E^{p}_{3} \E^{q}_{1}} 
 \ln \biggl| 
  \frac{(m_2^2 - m_3^2 - m_1^2 + 2 p q)^2 - 4 (\E^{p}_{3})^2 (\E^{q}_{1})^2 }
       {(m_2^2 - m_3^2 - m_1^2 - 2 p q)^2 - 4 (\E^{p}_{3})^2 (\E^{q}_{1})^2 }
 \biggr|
 \;, \hspace*{6mm} \nn 
\ea
where $\E^p_i \equiv \sqrt{p^2 + m_i^2}$ and $B^{(0)}_{ }$
is from \eq\nr{B_reg}. For finite masses 
a representation of the undefined functions
reads~\cite{cclr}\footnote{%
 The function $L$ is singular in certain limits, for instance
 $L(m^2,\epsilon^2,\epsilon^2) = -\pi^2/6 - 2 \ln^2(m/\epsilon)$
 for $\epsilon\to 0^+$.
 }
\ba
 R(m_1^2,m_2^2,m_3^2) & = & 
 \sqrt{m_1^4 + m_2^4 + m_3^4 - 2 m_1^2 m_2^2 - 2 m_1^2 m_3^2 - 2 m_2^2 m_3^2}
 \;, \la{R} \\ 
 L(m_1^2,m_2^2,m_3^2) & = & 
 \mbox{Li}^{ }_2 \Bigl( - \frac{t^{ }_3 m^{ }_2}{m^{ }_1} \Bigr)
  + 
 \mbox{Li}^{ }_2 \Bigl( - \frac{t^{ }_3 m^{ }_1}{m^{ }_2} \Bigr)
  + \frac{\pi^2}{6} 
  + \frac{\ln^2 t^{ }_3}{2}
 \nn 
 & & \; + \, 
 \fr12 \biggl[ 
   \ln\Bigl( t^{ }_3 + \frac{m^{ }_2}{m^{ }_1} \Bigr)
  - 
   \ln\Bigl( t^{ }_3 + \frac{m^{ }_1}{m^{ }_2} \Bigr)
  + \fr34 \ln \Bigl( \frac{m_1^2}{m_2^2} \Bigr) 
 \biggr] \ln\Bigl( \frac{m_1^2}{m_2^2} \Bigr)
 \;, \hspace*{4mm} \la{L} \\ 
 t^{ }_3 & = & 
 \frac{m_3^2 - m_1^2 - m_2^2 + R(m_1^2,m_2^2,m_3^2)}{2 m^{ }_1 m^{ }_2}
 \;. 
\ea 
The functions $H^{(n)}_{ }$ are related through
$
 H^{(n)}_{ }(\{m^{ }_i\}) = 
 \fr12 \bmu^2\, \partial H^{(n+1)}_{ }(\{m^{ }_i\}) /
 \partial \bmu^2
$.

Given that the function $H$ is always multiplied by $\sim g^2 m^2$ in the 
effective potential, the order $\sim T^2$ in the high-$T$
is sufficient for reaching the order $\sim g^2 m^2 T^2$ for $V^{ }_2$. 
To this accuracy (cf.\ e.g.\ refs.~\cite{az1,az2}),
\ba
 H(m^{ }_{b1},m^{ }_{b2},m^{ }_{b3}) & = & 
 \frac{T^2}{(4\pi)^2}
 \biggl(
   \frac{1}{4\epsilon}
  + \ln \frac{\bmu}{m^{ }_{b1} + m^{ }_{b2} + m^{ }_{b3} } + \fr12  
 \biggr) + \rmO\Bigl(\frac{\epsilon T^2}{\pi^2}, 
 \frac{ m^{ }_{bi} T}{\pi^3}\Bigr) 
     \;, \hspace*{8mm} \\ 
 H(m^{ }_{b1},m^{ }_\rmi{\sl f2},m^{ }_\rmi{\sl f3}) & = & 
 \rmO\Bigl(\frac{\epsilon T^2}{\pi^2},
 \frac{ m^{ }_{b1} T}{\pi^3}\Bigr)
     \;. 
\ea

%
\subsection{Function $H(\underline{m^{ }_1}, m^{ }_2, m^{ }_3)$}
\la{ss:Hbar}

The final ingredient needed is a variant of the ``sunset'' sum-integral
in \eq\nr{H}, with one line weighted by an additional spatial momentum:  
\be
 H(\underline{m^{ }_1}, m^{ }_2, m^{ }_3)
 \; \equiv \; 
 \Tint{P,Q} \frac{p^2}{(P^2 + m_1^2)(Q^2 + m_2^2) [(P+Q)^2 + m_3^2]}
 \;. \la{Hbar}
\ee
The Matsubara sums can be carried out as before, and for the vacuum
parts Lorentz symmetry allows furthermore to write
\be
 H^{ }_0(\underline{m^{ }_1}, m^{ }_2, m^{ }_3)
 = 
 \frac{D-1}{D}
 \Bigl[ 
   - m_1^2 
     H^{ }_0(m^{ }_1, m^{ }_2, m^{ }_3) + 
     I^{ }_0(m^{ }_2) I^{ }_0(m^{ }_3) 
 \Bigr]
 \;. 
\ee 
Expressing the result like in \eq\nr{H_expand} 
and omitting an overall $\mu^{-4\epsilon}$, we get
\ba
 && \hspace*{-1.0cm}
 H^{(-2)}_{ }(\underline{m^{ }_1}, m^{ }_2, m^{ }_3)
 \nn & = & 
  \frac{3\bigl[m_1^2( m_1^2 + m_2^2 + m_3^2) + 2\, m_2^2 m_3^2 \bigr]}
 {8(4\pi)^4}
 \;, \\[2mm] 
 && \hspace*{-1.0cm}
 H^{(-1)}_{ }(\underline{m^{ }_1}, m^{ }_2, m^{ }_3)
 \nn & = & 
  \frac{17 m_1^2( m_1^2 + m_2^2 + m_3^2) + 22\, m_2^2 m_3^2 }
  {16(4\pi)^4}
 \nn & + & 
  \frac{3}{4(4\pi)^4}
  \biggl[ 
    m_1^4 \ln\Bigl( \frac{\bmu^2}{m_1^2} \Bigr)
   + m_2^2 (m_1^2 + m_3^2) \ln\Bigl( \frac{\bmu^2}{m_2^2} \Bigr)
   + m_3^2 (m_1^2 + m_2^2) \ln\Bigl( \frac{\bmu^2}{m_3^2} \Bigr)
  \biggr] 
 \nn & + & 
 \frac{1}{(4\pi)^2}
 \biggl\{ 
     I^{(0)}_\rmii{$T$}(\underline{m^{ }_1}) 
   + \fr13\Bigl[ 
 I^{(0)}_\rmii{$T$}(\underline{m^{ }_2}) + 
 I^{(0)}_\rmii{$T$}(\underline{m^{ }_3})
      \Bigr]
 \biggr\} 
 \nn & + &   
 \frac{1}{4(4\pi)^2}
 \Bigl[ 
 (m_2^2 - 3 m_1^2 - 3 m_3^2)\, I^{(0)}_\rmii{$T$}(m^{ }_2)
 + 
 (m_3^2 - 3 m_1^2 - 3 m_2^2)\, I^{(0)}_\rmii{$T$}(m^{ }_3)
 \biggr]
 \;, \\[2mm] 
 && \hspace*{-1.0cm}
  H^{(0)}_{ }( \underline{m^{ }_1}, m^{ }_2, m^{ }_3 )
 \nn & = & 
  \frac{77\bigl[m_1^2( m_1^2 + m_2^2 + m_3^2) + 62\, m_2^2 m_3^2 \bigr]}
  {32(4\pi)^4}
  + 
  \frac{\pi^2\bigl[m_1^2( m_1^2 + m_2^2 + m_3^2) + 2\, m_2^2 m_3^2 \bigr]}
  {16(4\pi)^4}
 \nn & + & 
  \frac{3}{8(4\pi)^4}
  \biggl[ 
    m_1^4 \ln^2\Bigl( \frac{\bmu^2}{m_1^2} \Bigr)
   + m_2^2 (m_1^2 + m_3^2) \ln^2 \Bigl( \frac{\bmu^2}{m_2^2} \Bigr)
   + m_3^2 (m_1^2 + m_2^2) \ln^2 \Bigl( \frac{\bmu^2}{m_3^2} \Bigr)
  \biggr] 
 \nn & + & 
  \frac{1}{8(4\pi)^4}
  \biggl[ 
    17 m_1^4 \ln\Bigl( \frac{\bmu^2}{m_1^2} \Bigr)
   + m_2^2 (17 m_1^2 + 11 m_3^2) \ln \Bigl( \frac{\bmu^2}{m_2^2} \Bigr)
   + m_3^2 (17 m_1^2 + 11 m_2^2) \ln \Bigl( \frac{\bmu^2}{m_3^2} \Bigr)
  \biggr] 
 \nn & + & 
 \frac{1}{(4\pi)^4}
 \biggl\{ 
 \frac{3 m_1^2}{8} 
 \biggl[ 
  (m_1^2 + m_2^2 - m_3^2)\,
    \ln\Bigl( \frac{\bmu^2}{m_1^2} \Bigr)
    \ln\Bigl( \frac{\bmu^2}{m_2^2} \Bigr)
 + (m_1^2 + m_3^2 - m_2^2 )\,
    \ln\Bigl( \frac{\bmu^2}{m_1^2} \Bigr)
    \ln\Bigl( \frac{\bmu^2}{m_3^2} \Bigr)
 \nn 
 & + & 
   (m_2^2 + m_3^2 - m_1^2 )\,
    \ln\Bigl( \frac{\bmu^2}{m_2^2} \Bigr)
    \ln\Bigl( \frac{\bmu^2}{m_3^2} \Bigr)
 \biggr] 
 \; + \;  
 \frac{3 m_2^2 m_3^2}{4}
    \ln\Bigl( \frac{\bmu^2}{m_2^2} \Bigr)
    \ln\Bigl( \frac{\bmu^2}{m_3^2} \Bigr)
 \nn & - & 
 \frac{3 m_1^2}{4}
  R(m_1^2,m_2^2,m_3^2)\, L(m_1^2,m_2^2,m_3^2) 
 \biggr\} 
 \nn 
 & + & 
 I^{(0)}_\rmii{$T$}(\underline{m^{ }_1})\, 
 \re B^{(0)}_{}(-i m^{ }_1;m^{ }_2, m^{ }_3)
 \nn 
 & + & 
 I^{(0)}_\rmii{$T$}(\underline{m^{ }_2})\, 
 \biggl\{ \frac{R^2(m_1^2,m_2^2,m_3^2) + 3 m_1^2 m_2^2}{3 m_2^4}
 \,\re B^{(0)}_{}(-i m^{ }_2;m^{ }_3, m^{ }_1)
 \nn 
 & + &  \frac{1}{18 m_2^4 (4\pi)^2}
  \biggl[
  -  6 m_1^2 (m_1^2 + m_2^2 - m_3^2) \ln \Bigl( \frac{\bmu^2}{m_1^2} \Bigr)
  +
    6 m_3^2 (m_1^2 + 2 m_2^2 - m_3^2) \ln \Bigl( \frac{\bmu^2}{m_3^2} \Bigr)
 \nn 
 & - & 6 m_1^4 + m_2^4 - 6 m_3^4 - 9 m_1^2 m_2^2
    + 12 m_1^2 m_3^2 + 9 m_2^2 m_3^2
  \biggr]
 \biggr\} \; + \; (2\leftrightarrow 3)
 \nn 
 & + & 
 I^{(0)}_\rmii{$T$}(m^{ }_2)\, 
 \biggl\{ \frac{R^2(m_1^2,m_2^2,m_3^2)}{4 m_2^2}
 \,\re B^{(0)}_{}(-i m^{ }_2;m^{ }_3, m^{ }_1)
 \nn 
 & -  &  \frac{1}{4 m_2^2 (4\pi)^2}
  \biggl[
     m_1^2 (m_1^2 + m_2^2 - m_3^2) \ln \Bigl( \frac{\bmu^2}{m_1^2} \Bigr)
  +
     m_3^2 (- m_1^2 + m_2^2 + m_3^2) \ln \Bigl( \frac{\bmu^2}{m_3^2} \Bigr)
 \nn 
 & + &  m_1^4 + m_3^4 + m_1^2 m_2^2
     - 2 m_1^2 m_3^2 +  m_2^2 m_3^2
  \biggr]
 \biggr\} \; + \; (2\leftrightarrow 3)
 \nn 
 & + & 
 \frac{1}{(4\pi)^2}
 \biggl\{ 
     I^{(1)}_\rmii{$T$}(\underline{m^{ }_1}) 
   + \fr13\Bigl[ 
 I^{(1)}_\rmii{$T$}(\underline{m^{ }_2}) + 
 I^{(1)}_\rmii{$T$}(\underline{m^{ }_3})
      \Bigr]
 \biggr\} 
 \nn & + &   
 \frac{1}{4(4\pi)^2}
 \Bigl[ 
 (m_2^2 - 3 m_1^2 - 3 m_3^2)\, I^{(1)}_\rmii{$T$}(m^{ }_2)
 + 
 (m_3^2 - 3 m_1^2 - 3 m_2^2)\, I^{(1)}_\rmii{$T$}(m^{ }_3)
 \biggr]
 \nn 
 & + & 
 \int_{0}^\infty \!  
 \frac{{\rm d}p\, {\rm d} q\, p^3 q\, n(\E^{p}_{1}) n(\E^{q}_{2})}
 {32\pi^4 \E^{p}_{1} \E^{q}_{2}} 
 \ln \biggl| 
  \frac{(m_3^2 - m_1^2 - m_2^2 + 2 p q)^2 - 4 (\E^{p}_{1})^2 (\E^{q}_{2})^2 }
       {(m_3^2 - m_1^2 - m_2^2 - 2 p q)^2 - 4 (\E^{p}_{1})^2 (\E^{q}_{2})^2 }
 \biggr| \; + \; (2\leftrightarrow 3)
 \nn[2mm] 
 & + & 
 \int_{0}^\infty \!  
 \frac{{\rm d}p\, {\rm d} q\, p^2 q^2\,  n(\E^{p}_{2}) n(\E^{q}_{3})}
 {4\pi^4  \E^{p}_{2} \E^{q}_{3}} 
 \biggl\{ 1 \; + \; \frac{ \E^{p}_{2} \E^{q}_{3}}{4p q} 
 \ln \biggl| 
  \frac{(m_1^2 - m_2^2 - m_3^2)^2 - 4 ( p q - \E^{p}_{2} \E^{q}_{3} )^2 }
       {(m_1^2 - m_2^2 - m_3^2)^2 - 4 ( p q + \E^{p}_{2} \E^{q}_{3} )^2 }
 \biggr|
 \nn[2mm] 
 & + & 
 \frac{ (\E^{p}_{2})^2 + (\E^{q}_{3})^2 - m_1^2 }{8p q}
 \ln \biggl| 
  \frac{(m_1^2 - m_2^2 - m_3^2 + 2 p q)^2 - 4 (\E^{p}_{2})^2 (\E^{q}_{3})^2 }
       {(m_1^2 - m_2^2 - m_3^2 - 2 p q)^2 - 4 (\E^{p}_{2})^2 (\E^{q}_{3})^2 }
 \biggr|
 \biggr\}
 \;, \la{Hbar_full} \hspace*{6mm}  
\ea
where $\E^p_i \equiv \sqrt{p^2 + m_i^2}$, $B^{(0)}_{ }$
is from \eq\nr{B_reg}, and $R$ and $L$ are from 
\eqs\nr{R} and \nr{L}, respectively. 
The functions are related through
$
 H^{(n)}_{ } 
 = 
 \fr12 \bmu^2\, \partial 
  H^{(n+1)}_{ } 
 /
 \partial \bmu^2
$.

The numerical evaluation of \eq\nr{Hbar_full} is straightforward
if all masses are of similar orders of magnitude. In contrast, if there
is a hierarchy between the masses (cf.\ e.g.\ \eq\nr{gff}), 
care must be taken in order to avoid 
significance loss in the numerics. For instance, the coefficient
multiplying $ I^{(0)}_\rmii{$T$}(\underline{m^{ }_2}) $ has a finite
limit for $m^{ }_2\to 0$, but many individual terms within the curly
brackets diverge as $\sim 1/m_2^4$. A similar problem appears in the 
coefficient 
multiplying $ I^{(0)}_\rmii{$T$}(m^{ }_2) $, even though
divergences are only $\sim 1/m_2^2$ in this case. 
It may also be noted
that the coefficients of 
$ I^{(0)}_\rmii{$T$}(\underline{m^{ }_2}) $ and 
$ I^{(0)}_\rmii{$T$}(m^{ }_2) $
have cusps at $m^{ }_2 = m^{ }_1 + m^{ }_3$, originating
from the function $B^{(0)}_{ }$, which cancel 
against corresponding cusps originating from the last three
rows of \eq\nr{Hbar_full}. For a proper cancellation of the
cusps, all terms involved
need to be evaluated with good precision. 
A powerful crosscheck on the 
numerics is provided by the
high-$T$ limit, which can be given in analytic form
(cf.\ \eqs\nr{Hbar_highT1}--\nr{Hbar_highT3}). 

In the high-$T$ limit, making use of relations determined 
in ref.~\cite{masters},\footnote{%
 We thank Y.~Schr\"oder for locating the necessary relations. 
 } 
we get
\ba
 H(\underline{m^{ }_{b1}},m^{ }_{b2},m^{ }_{b3}) & = & 
 \frac{T^4}{72}
 \biggl[
  \frac{1}{4\epsilon} + 
  \ln\biggl(\frac{\bmu e^{\gammaE}}{2 T} \biggr) - \frac{\zeta'(2)}{\zeta(2)}
 \biggr]
 \la{Hbar_highT1} \\ & + & 
 \frac{D-1}{2} 
 I(\zerob) \, 
 \bigl[I^{ }_{n=0}(m^{ }_{b2}) + I^{ }_{n=0}(m^{ }_{b3}) \bigr]
 \;  + \; \frac{m^{ }_{2b}m^{ }_{3b} T^2 }{(4\pi)^2}
 \nn & - & 
 \frac{m_{1b}^2 T^2 }{(4\pi)^2}
 \biggl[
   \frac{1}{4\epsilon}
  + \ln  \biggl(
  \frac{\bmu}{m^{ }_{b1} + m^{ }_{b2} + m^{ }_{b3} } \biggr) + \fr12  
 \biggr]
 \; + \; \rmO\Bigl(
 \epsilon T^4,
 \frac{ m^{2}_{ } T^2 }{\pi^2}\Bigr) 
     \;, \nn[2mm]  
 H(\underline{m^{ }_{b1}},m^{ }_\rmi{\sl f2},m^{ }_\rmi{\sl f3}) & = & 
 \frac{T^4}{288}
 \biggl[
  \frac{1}{4\epsilon} + 
  \ln\biggl(\frac{\bmu e^{\gammaE}}{4T} \biggr)
  - \frac{\zeta'(2)}{\zeta(2)}
 \biggr]
 \; + \; 
 \rmO\Bigl(
 \epsilon T^4,
 \frac{ m^{2}_{ } T^2}{\pi^2}\Bigr)
     \;, \\[2mm] 
 H(\underline{m^{ }_\rmi{\sl f1}},m^{ }_\rmi{\sl f2},m^{ }_{b3}) & = & 
 -\frac{T^4}{144}
 \biggl[
  \frac{1}{4\epsilon} + 
  \ln\biggl(\frac{\bmu e^{\gammaE}}{T} \biggr)
  - \frac{3\ln(2)}{2} - \frac{\zeta'(2)}{\zeta(2)}
 \biggr]
 \la{Hbar_highT3} \\ 
 & + & 
 \frac{D-1}{2} 
 I(\zerof)\, I^{ }_{n=0}(m^{ }_{b3})
 \; + \; 
 \rmO\Bigl( \epsilon T^4,
 \frac{ m^{2}_{ } T^2 }{\pi^2} \Bigr)
     \;. \nonumber 
\ea
Here the so-called linear terms, of the type $\rmO(m T^3)$, have
been written in a $D$-dimensional form, permitting for a crosscheck
of their $D$-dimensional cancellation in the full result
(cf.\ the discussion around the end of \se\ref{se:resum}). 

The function $H(\underline{m^{ }_1},m^{ }_2,m^{ }_3)$
always appears in a difference containing various masses, so 
that the leading
term $\propto T^4$ of the high-$T$ expansion
drops out from the effective potential. Moreover, 
at $\rmO(m^2)$, only the {\em non-analytic terms} originating from 
Matsubara zero modes have been kept in the expressions above, 
given that analytic terms lead to $v$-independent 
structures of the type $\sim g^2 (\mWt^2-\mW^2)T^2 \sim g^4 T^4$. 

%
\section{2-loop diagrams}
\la{se:diags}

In order to list all contributions to $V^{ }_2$, we make use of the
master sum-integrals defined in appendices~\ref{ss:I}--\ref{ss:Hbar}. 
The diagrams are of three types, illustrated in \fig\ref{fig:graphs}.
We denote the particles circling
in the loops by scalar (s), vector (v), ghost (g), or fermion (f). 
Then the various contributions read 
(some $v$-independent terms have been dropped for simplicity)
\ba
 \mbox{(sss)} & = & 
 -3 v^2 \lambda_1^2 \bigl[H(\mht,\mht,\mht) + H(\mht,\mGt,\mGt) \bigr]
 \nn & - & 
 \frac{v^2}{4}\bigl[ (\lambda^{ }_3 + \lambda^{ }_4 + \lambda^{ }_5)^2 
  H(\mht,\mHt,\mHt) + (\lambda^{ }_3 + \lambda^{ }_4 - \lambda^{ }_5)^2 
  H(\mht,\mAt,\mAt) \bigr] \hspace*{10mm}
 \nn & - & 
 \frac{v^2}{4}\bigl[ (\lambda^{ }_4 + \lambda^{ }_5)^2 
  H(\mGt,\mHt,\mHpmt) +  (\lambda^{ }_4 - \lambda^{ }_5)^2 
  H(\mGt,\mAt,\mHpmt) \bigr]
 \nn & - & 
 \frac{v^2}{2} \bigl[ \lambda^{2}_3\, 
  H(\mht,\mHpmt,\mHpmt) + \lambda^{2}_5\, 
  H(\mGt,\mHt,\mAt) \bigr]
 \;, \\ 
 \mbox{(ss)} & = & 
 \frac{3\lambda^{ }_1}{4}
 \bigl[
  I^2(\mht) + 2 I(\mht) I(\mGt) + 5 I^2(\mGt)  
 \bigr]
 \nn & + &
 \frac{\lambda^{ }_2}{2}
 \bigl[
    I^2(\mHt) + I^2(\mAt) + 2 I^2(\mHpmt)
 \bigr] + 
 \frac{\lambda^{ }_2}{4}
 \bigl[ I(\mHt) + I(\mAt) + 2 I(\mHpmt)  \bigr]^2
 \nn & + &
 \frac{\lambda^{ }_3 + \lambda^{ }_4 + \lambda^{ }_5}{4}
 \bigl[ I(\mht) I(\mHt) + I(\mGt) I(\mAt) \bigr]
 \nn & + &
 \frac{\lambda^{ }_3 + \lambda^{ }_4 - \lambda^{ }_5}{4}
 \bigl[ I(\mht) I(\mAt) + I(\mGt) I(\mHt) \bigr]
 \; + \; 
  (\lambda^{ }_3 + \lambda^{ }_4)
  I(\mGt) I(\mHpmt)
 \nn & + &
 \frac{\lambda^{ }_3}{2}
 \bigl\{ I(\mht) I(\mHpmt) + I(\mGt)\bigl[ 
  I(\mHt) + I(\mAt) + I(\mHpmt) \bigr] \bigr\}
 \;, \\ 
 \mbox{(s)} & = & 
   \frac{\delta \mh^2 - \delta \mphiT^2
  -\mht^2 \delta Z^{ }_\phi }{2} \, I(\mht) 
 + \frac{3 (\delta \mG^2 - \delta \mphiT^2
  -\mGt^2 \delta Z^{ }_\phi) }{2} \, I(\mGt)
 \nn & + & 
 \frac{\delta \mH^2
  - \delta \mchiT^2 -\mHt^2 \delta Z^{ }_\chi }{2} \, I(\mHt) 
 + \frac{\delta \mA^2
  - \delta \mchiT^2 -\mAt^2 \delta Z^{ }_\chi }{2} \, I(\mAt) 
 \nn 
 & + & 
 \bigl(\delta \mHpm^2
   - \delta \mchiT^2 -\mHpmt^2 \delta Z^{ }_\chi \bigr) \, I(\mHpmt)  
 \;, \la{s} \\ 
 \mbox{(ssv)} & = & 
 -\frac{3 g_2^2}{8}
 \Bigl\{ (\mWt^2 - 2 \mht^2 - 2 \mGt^2) H(\mWt,\mht,\mGt)
         + (\mWt^2 - 4 \mGt^2) H(\mWt,\mGt,\mGt)
 \nn  &  - & \bigl[ I(\mht) + I(\mGt) \bigr] I(\mGt)
         + 2 \bigl[ I(\mht) + 3 I(\mGt) \bigr] I(\mWt)
 \nn & + & 
  H(\underline{\mWt},\mht,\mGt)
  -2 H(\mWt,\underline{\mht},\mGt)
  -2 H(\mWt,\mht,\underline{\mGt})
 \nn & + & 
  H(\underline{\mWt},\mGt,\mGt)
  -4 H(\mWt,\underline{\mGt},\mGt)
 \nn & - & 
  H(\underline{\mW},\mht,\mGt)
  + 2 H(\mW,\underline{\mht},\mGt)
  + 2 H(\mW,\mht,\underline{\mGt})
 \nn & - & 
   H(\underline{\mW},\mGt,\mGt)
  + 4 H(\mW,\underline{\mGt},\mGt)
 \Bigr\} 
 \nn & - & 
 \frac{g_2^2}{8}
 \Bigl\{ (\mWt^2 - 2 \mHt^2 - 2 \mAt^2) H(\mWt,\mHt,\mAt)
 \; + \; (\mWt^2 - 4 \mHpmt^2) H(\mWt,\mHpmt,\mHpmt)
 \nn & + & 2 (\mWt^2 - 2 \mHt^2 - 2 \mHpmt^2) H(\mWt,\mHt,\mHpmt)
 \nn & + & 
  2 (\mWt^2 - 2 \mAt^2 - 2 \mHpmt^2) H(\mWt,\mAt,\mHpmt)
 \nn  &  - &  I(\mHt) I(\mAt) - 
 \bigl[2 I(\mHt) + 2 I(\mAt) + I(\mHpmt) \bigr] I(\mHpmt)
  \nn  &  + & 6 \bigl[ I(\mHt) + I(\mAt) + 2 I(\mHpmt) \bigr] I(\mWt)
 \nn & + & 
  H(\underline{\mWt},\mHt,\mAt)
  -2 H(\mWt,\underline{\mHt},\mAt)
  -2 H(\mWt,\mHt,\underline{\mAt})
 \nn & + & 
  2 H(\underline{\mWt},\mHt,\mHpmt)
  -4 H(\mWt,\underline{\mHt},\mHpmt)
  -4 H(\mWt,\mHt,\underline{\mHpmt})
 \nn & + & 
  2 H(\underline{\mWt},\mAt,\mHpmt)
  -4 H(\mWt,\underline{\mAt},\mHpmt)
  -4 H(\mWt,\mAt,\underline{\mHpmt})
 \nn & + & 
  H(\underline{\mWt},\mHpmt,\mHpmt)
  -4 H(\mWt,\underline{\mHpmt},\mHpmt)
 \nn & - & 
  H(\underline{\mW},\mHt,\mAt)
  +2 H(\mW,\underline{\mHt},\mAt)
  +2 H(\mW,\mHt,\underline{\mAt})
 \nn & - & 
  2 H(\underline{\mW},\mHt,\mHpmt)
  +4 H(\mW,\underline{\mHt},\mHpmt)
  +4 H(\mW,\mHt,\underline{\mHpmt})
 \nn & - & 
  2 H(\underline{\mW},\mAt,\mHpmt)
  +4 H(\mW,\underline{\mAt},\mHpmt)
  +4 H(\mW,\mAt,\underline{\mHpmt})
 \nn & - & 
  H(\underline{\mW},\mHpmt,\mHpmt)
  +4 H(\mW,\underline{\mHpmt},\mHpmt)
 \Bigr\} 
 \;, \\ 
 \mbox{(sv)} & = & 
 \frac{3 g_2^2}{8}
 \bigl[ I(\mht) + 3 I(\mGt) \bigr]
 \bigl[ I(\mWt) + (D-1) I(\mW) \bigr]
 \nn & + & 
  \frac{3 g_2^2}{8}
 \bigl[ I(\mHt) + I(\mAt) + 2 I(\mHpmt) \bigr]
 \bigl[ I(\mWt) + (D-1) I(\mW) \bigr]
 \;, \\ 
 \mbox{(svv)} & = & 
 -\frac{3 g_2^2 \mW^2}{4}
 \bigl[
   H(\mht,\mWt,\mWt) + (D-1) H(\mht,\mW,\mW)  
 \bigr]
 \;, \\ 
 \mbox{(sgg)} & = & 
 \frac{3 g_2^2 \mW^2}{8}
 \bigl[
   H(\mht,\mW,\mW) - 2 H(\mGt,\mW,\mW)  
 \bigr]
 \;, \\ 
 \mbox{(vvv)} & = & 
 -\frac{3 g_2^2}{2}
 \Bigl\{ 
 (D-1) (\mWt^2 -  4\mW^2) H(\mWt,\mW,\mW)
 \nn & + & (D-1) \bigl[ 4  I(\mWt)  - I(\mW)  \bigr] I(\mW)
 \nn & + & (D-1) \bigl[ H(\underline{\mWt},\mW,\mW)
  - 4 H(\mWt,\underline{\mW},\mW) + 3 H(\underline{\mW},\mW,\mW) \bigr]
 \nn & - & H(\underline{\mW},\mWt,\mWt) 
  + 4  H(\mW,\underline{\mWt},\mWt)
  - 3  H(\underline{\mW},\mW,\mW)
 \Bigr\} 
 \;, \\ 
 \mbox{(ggv)} & = & 
 -\frac{3 g_2^2}{2}
 \Bigl\{ 
 (2\mW^2 -  \mWt^2) H(\mWt,\mW,\mW)
         +  \bigl[ I(\mW) - 2 I(\mWt) \bigr] I(\mW)
 \nn & - &  H(\underline{\mWt},\mW,\mW)
  + 2 H(\mWt,\underline{\mW},\mW)
  - H(\underline{\mW},\mW,\mW)
 \Bigr\} 
 \;, \\ 
 \mbox{(vv)} & = & 
 \frac{3 g_2^2}{2}
 (D-1) \bigl[ 2  I(\mWt)  +(D-2) I(\mW)  \bigr] I(\mW)
 \;, \\ 
 \mbox{(v)} & = & 
 \fr32\Bigl\{
   \delta Z^{ }_\xi\, 
   \bigl[  I(\underline{\mWt}) -  I(\underline{\mW}) \bigr]
 \;  + \; (D-1) \mW^2 (\delta Z^{ }_{g^2} + \delta Z^{ }_\phi)
 \, I(\mW)
 \nn &  + & \bigl[ \mWt^2 \delta Z^{ }_\xi
  - \mE^2 \delta Z^{ }_A  + 
   \mW^2 (\delta Z^{ }_{g^2} + \delta Z^{ }_\phi) \bigr] \, I(\mWt)
  \Bigr\} 
 \; - \; \fr32 \mE^2 I(\mWt) 
 \;, \la{v} \\ 
 \mbox{(g)} & = & 
 - 3 
 \mW^2\biggl(\delta Z^{ }_\xi + \delta Z^{ }_{g^2}
  + \frac{ \delta Z^{ }_\phi + \delta Z^{ }_v }{2} \biggr) \, I(\mW)
 \;,  \\ 
 \mbox{(sff)} & = & 
 - \frac{3 h_t^2}{2}
 \Bigl\{ 
  (\mht^2 - 4 \mt^2) H(\mht,\mt^{ } ,\mt^{ } )
  + \mGt^2 H(\mGt,\mt^{ } ,\mt^{ } )
 \nn & + & 2 (\mGt^2 - \mt^2) H(\mGt,\mt^{ } ,\zerof)
 -2 \bigl[ I(\mt^{ } ) + I(\zerof) \bigr] I(\mt^{ } )
 \nn & + & 
  2 I(\mht) I(\mt^{ } )
 + 2 I(\mGt) \bigl[ I(\zerof) + 2 I(\mt^{ } ) \bigr]
 \Bigr\}
 \;, \\
 \mbox{(gff)} & = & 
 \frac{3 g_2^2}{8} \Bigl\{ 
 (\mWt^2 - 2 \mt^2) H(\mWt,\mt^{ } ,\mt^{ } )
 - (D-1) (\mW^2 - 2 \mt^2 ) H(\mW,\mt^{ } ,\mt^{ } )
 \nn & + & 
 (D-2) I^2(\mt^{ } ) 
 + 2 \bigl[ I(\mWt)
 -  (D-1) I(\mW) \bigr] I(\mt^{ } )
 \nn & - &
 2 H(\underline{\mW},\mt^{ } ,\mt^{ } ) 
 + 4 H(\mW,\underline{\mt^{ } },\mt^{ } ) 
 \; + \; 
 2 H(\underline{\mWt},\mt^{ } ,\mt^{ } ) 
 - 4 H(\mWt,\underline{\mt^{ } },\mt^{ } ) 
 \Bigr\}
 \nn & + & 
 \frac{3 g_2^2}{2} \Bigl\{ 
 (\mWt^2 - \mt^2) H(\mWt,\mt^{ } ,\zerof)
 - (D-1) (\mW^2 - \mt^2 ) H(\mW,\mt^{ } ,\zerof)
 \nn & + & 
 (D-2) I(\mt^{ } )I(\zerof) + 
 \bigl[ I(\mWt) 
 - (D-1) I(\mW) \bigr]\bigl[ I(\mt^{ } ) + I(\zerof) \bigr]
 \nn & - &
 2 H(\underline{\mW},\mt^{ } ,\zerof) 
 + 2 H(\mW,\underline{\mt^{ } },\zerof) 
 + 2 H(\mW,\mt^{ } ,\underline{\zerof}) 
 \nn & + & 
 2 H(\underline{\mWt},\mt^{ } ,\zerof) 
 - 2 H(\mWt,\underline{\mt^{ } },\zerof) 
 - 2 H(\mWt,\mt^{ } ,\underline{\zerof}) 
 \Bigr\}
 \nn & + & 
 \frac{3 (8 \nG - 5) g_2^2}{8} \Bigl\{ 
  \mWt^2 H(\mWt,\zerof,\zerof)
 - (D-1) \mW^2 H(\mW,\zerof,\zerof)
 \nn & + & 
 (D-2) I^2(\zerof) 
 + 2 \bigl[ I(\mWt) - (D-1) I(\mW) \bigr]\,  I(\zerof)
 \nn & - &
 2 H(\underline{\mW},\zerof,\zerof) 
 + 4 H(\mW,\underline{\zerof},\zerof) 
 \; + \; 
 2 H(\underline{\mWt},\zerof,\zerof) 
 - 4 H(\mWt,\underline{\zerof},\zerof) 
 \Bigr\}
 \nn & + & 
 {4 g_3^2} \Bigl\{ 
 (\mC^2 - 4 \mt^2) H(\mC,\mt^{ } ,\mt^{ } )
 \nn & + & 
 (D-2) I^2(\mt^{ } ) 
 + 2 \bigl[ I(\mC)
 -  (D-1) I(\zerob) \bigr] I(\mt^{ } )
 \nn & - &
 2 H(\underline{\zerob},\mt^{ } ,\mt^{ } ) 
 + 4 H(\zerob,\underline{\mt^{ } },\mt^{ } ) 
  \nn &  + & 
 2 H(\underline{\mC},\mt^{ } ,\mt^{ } ) 
 - 4 H(\mC,\underline{\mt^{ } },\mt^{ } ) 
 \Bigr\}
 \;,  \la{gff} \\
 \mbox{(f)} & = & 
 -6 \mt^2 \bigl( \delta Z^{ }_\phi + \delta Z^{ }_{h_t^2}\bigr) \, I(\mt^{ } )
 \;. 
\ea

%
\section{Vacuum renormalization}
\la{se:vac}

For the thermal computations, we need to know the running couplings
as functions of the $\msbar$ scale $\bmu$
up to a scale $\bmu \sim \pi T$, cf.\ \eq\nr{scale}. These can be obtained from
renormalization group equations, provided that the initial values 
are known at some scale $\bmu\sim \mZ$. In order to obtain the latter, 
we compute physical pole masses and the Fermi constant in terms of the
$\msbar$ parameters, and then invert the relations in order to express
the $\msbar$ couplings at $\bmu = \mZ$ 
in terms of the physical ones. For the Standard
Model, these relations were determined up to 1-loop level in 
ref.~\cite{generic},\footnote{%
 In eq.~(193) of ref.~\cite{generic}, there is a term
 $-\frac{8}{3}t^2\ln h$ missing from within the square brackets.  
 }
and here we extend the relations to the IDM. Closely related
expressions for the IDM can be found in ref.~\cite{pole}. 

%
\subsection{Basis functions}

In order to display the results for physical quantities, we make
use of standard Passarino-Veltman type functions, which we have 
defined in Euclidean spacetime:  
\ba
     A^{ }_{ }(m) & \equiv & 
     \int^{ }_{P} \frac{1}
     {P^2 + m^2}
     \;, \la{A} \\
     B^{ }_{ }(K;m_1^{ },m_2^{ }) & \equiv & 
     \int^{ }_{P} \frac{1}
     {[(P+K)^2+m_1^2](P^2+m_2^2)}
     \;,  \\ 
     K^{ }_\mu C^{ }_{ }(K;m_1^{ },m_2^{ }) & \equiv & 
     \int^{ }_{P} \frac{P^{ }_\mu}
     {[(P+K)^2+m_1^2](P^2+m_2^2)}
     \;, \\
     C^{ }_{ }(K;m_1^{ },m_2^{ })
     & = & 
     \frac{1}{2 K^2} 
      \bigl[ A(m^{ }_2) - A(m^{ }_1)
          - (K^2 + m_1^2 - m_2^2)\, B^{ }_{ }(K;m_1^{ },m_2^{ })
      \bigr]
     \;, \la{Cm}  \\ 
     D^{ }_{\mu\nu}(K;m_1^{ },m_2^{ }) & \equiv & 
     \int^{ }_{P} \frac{P^{ }_\mu P^{ }_\nu}
     {[(P+K)^2+m_1^2](P^2+m_2^2)}
     \la{Dmn} \\
     & = & 
     \frac{\delta^{ }_{\mu\nu} - \frac{D\, K^{ }_\mu K^{ }_\nu}{K^2}}
     {4(D-1)K^2}
     \Bigl\{ (K^2 - m_1^2 + m_2^2)\, A(m^{ }_1)
           +(K^2 + m_1^2 - m_2^2)\, A(m^{ }_2)
     \nn & & \;
           -\, \bigl[ K^4 + 2 K^2(m_1^2+m_2^2)
               + (m_1^2 - m_2^2)^2 \bigr] B(K;m^{ }_1,m^{ }_2)
     \Bigr\}
     \nn 
     & + & \frac{K^{ }_\mu K^{ }_\nu}{K^2}
     \bigl[ A(m^{ }_1) - m_2^2\, B(K;m^{ }_1,m^{ }_2) \bigr]
     \;. 
\ea
For $D = 4 -2\epsilon$ and writing 
$A = \sum_{n=-1}^{\infty} A^{(n)} \epsilon^n$ etc, 
the divergent parts of these functions can be expressed as 
\ba
 A^{(-1)}_{ }(m) & = & \frac{\mu^{-2\epsilon}}{(4\pi)^2}
 \bigl(-m^2\bigr)
 \;, \la{A_div} \\ 
 B^{(-1)}_{ }(K;m^{ }_1,m^{ }_2) & = & 
 \frac{\mu^{-2\epsilon}}{(4\pi)^2}
 \;, \la{B_div} \\ 
 C^{(-1)}_{ }(K;m^{ }_1,m^{ }_2) & = & 
 \frac{\mu^{-2\epsilon}}{(4\pi)^2} 
 \biggl( -\frac{1}{2} \biggr)
 \;, \la{C_div} \\ 
 D^{(-1)}_{\mu\nu}(K;m^{ }_1,m^{ }_2) & = & 
 \frac{\mu^{-2\epsilon}}{(4\pi)^2} 
 \biggl( 
  -  \frac{K^2 + 3 m_1^2 + 3 m_2^2}{12} \, \delta^{ }_{\mu\nu} 
  + \frac{K^{ }_\mu K^{ }_\nu}{3} 
 \biggr)
 \;. \la{D_div}
\ea
The finite parts of $A$ and $B$ read 
\ba
 A^{(0)}_{ }(m) & = & - \frac{m^2}{(4\pi)^2}
 \biggl( \ln\frac{\bmu^2}{m^2} + 1 \biggr)
 \;, \la{A_reg} \\ 
 B^{(0)}_{ }(K;m^{ }_1,m^{ }_2) & = & 
 \frac{1}{(4\pi)^2}
 \biggl[ 
   \ln\frac{\bmu^2}{m^{ }_1 m^{ }_2} + 2 + \frac{m_1^2 - m_2^2}{K^2}
   \ln\frac{m^{ }_1}{m^{ }_2}
 \la{B_reg} \\ 
 &  & \hspace*{-2cm} -\, \frac{2\sqrt{(m^{ }_1 - m^{ }_2)^2 +K^2}
          \sqrt{(m^{ }_1 + m^{ }_2)^2 +K^2} }{K^2}
   \,\mbox{artanh}\biggl( \frac{\sqrt{(m^{ }_1 - m^{ }_2)^2 +K^2}}
                      {\sqrt{(m^{ }_1 + m^{ }_2)^2 +K^2}} \biggr)
 \, \biggr]
 \;. \nonumber
\ea
Given that $B^{(0)}_{ }$ is a function of $K^2$ only, we use an 
implicit notation in which $K$ may denote either 
a vector or its modulus. 
The corresponding expressions after going to Minkowskian signature, 
i.e.\ $K \to -i \mathcal{K}$, 
are conventionally expressed in terms of a function $F$ defined by
\be
 B^{(0)}_{ }(-i\mathcal{K};m^{ }_1,m^{ }_2)  \;\equiv\; 
 \frac{1}{(4\pi)^2}
 \biggl[ 
   \ln\frac{\bmu^2}{m^{ }_1 m^{ }_2} 
   + 1 - \frac{m_1^2 + m_2^2}{m_1^2 - m_2^2}
   \ln\biggl( \frac{m^{ }_1}{m^{ }_2} \biggr)
   + F \Bigl( \frac{m^{ }_1}{\mathcal{K}},
              \frac{m^{ }_2}{\mathcal{K}} \Bigr)
 \biggr]
 \;. \la{F_def}
\ee
The (real part of) $F$ is given in \eq\nr{F} below.

%
\subsection{Gauge coupling renormalization}

The first quantity needed is the initial value of 
the SU$^{ }_\rmii{L}$(2) gauge 
coupling $g_2^2$ at the scale $\bmu\sim \mZ$. It can be expressed
in terms of the Fermi constant.
Including the contribution
of the new scalar degrees of freedom through a function~$\Delta$
(cf.\ \eq\nr{Delta}) we get
\ba
 g^2(\bmu) & = & g^2_0\biggl\{1+
 \frac{g^2_0}{16 \pi^2} 
 \biggl[\biggl(\frac{4 \nG}{3}-7\biggr)
 \ln\frac{\bmu^2}{\mW^2}
 + \Delta\Bigl( \frac{\mHpm}{\mW},\frac{\mH}{\mW} \Bigr)
 + \Delta\Bigl( \frac{\mHpm}{\mW},\frac{\mA}{\mW} \Bigr)
 \nonumber\\
 & - & \frac{33}{4} F(1,1)+
 \frac{1}{12}(h^4-4h^2+12)\re F(1,h) -  \frac{1}{2}(t^4+t^2-2)\re F(t,0) 
 \nn & - & 
 2\ln{t}-\frac{h^2}{24} +\frac{t^2}{4} +\frac{20 \nG}{9}-
 \frac{257}{72}\biggr] \biggr\} 
 \;, \la{g2}
\ea
where $g_0^2 \equiv 4\sqrt{2} G_\rmii{F}^{ } \mW^2$, 
$G_\rmii{F}^{ } = 1.166379\times 10^{-5}$ GeV$^{-2}$ is the Fermi constant, 
$h\equiv \mh/\mW$, $t\equiv \mt/\mW$, 
the masses $\mW,\mt,\mh,\mH,\mA$ and $\mHpm$ 
are the physical (vacuum) masses, 
and we have defined
\ba
 \Delta(r^{ }_1, r^{ }_2) & \equiv & 
 \frac{5}{36} - \frac{r_1^2+r_2^2}{24} - \frac{\ln(r^{ }_1 r^{ }_2)}{12}
 + 
 \frac{2 r_1^2  r_2^2  - r_1^2 - r_2^2
 }{12(r_1^2 - r_2^2)} \ln\biggl( \frac{r^{ }_1}{r^{ }_2} \biggr)
 \nn & + & 
 \frac{(r_1^2-r_2^2)^2-2(r_1^2 + r_2^2)+1}{12}\, \re F(r^{ }_1,r^{ }_2)
 \;. \la{Delta}
\ea
Here the function $F$, defined in \eq\nr{F_def}, has the real part
\ba
 \re F(r^{ }_1,r^{ }_2) & = & 
 1 + \biggl(\frac{r_1^2+r_2^2}{r_1^2 - r_2^2} + r_2^2 - r_1^2 \biggr)
 \ln\biggl( \frac{r^{ }_1}{r^{ }_2} \biggr)
 \nn & - & 2 
 \re\biggl[ 
  \sqrt{1 - (r^{ }_1 - r^{ }_2)^2}
  \sqrt{(r^{ }_1 + r^{ }_2)^2 - 1}
  \arctan\frac{\sqrt{1 - (r^{ }_1 - r^{ }_2)^2}}
 {  \sqrt{(r^{ }_1 + r^{ }_2)^2 - 1} } 
 \biggr]
  \;, \hspace*{7mm} \la{F}
\ea
with the special limits
\ba 
 F(1,1) & = & 2-\frac{\pi}{\sqrt{3}} \;, \\ 
 \re F(r,0) & = & 1+(r^2-1)\ln\biggl(1-\frac{1}{r^2}\biggr)
 \;, \quad r \ge 1
 \;.
\ea

%
\subsection{Pole masses and scalar coupling renormalizations}
\la{ss:couplingsmZ}

The other couplings can be expressed in terms of pole masses. 
For this purpose we compute the full 
renormalized on-shell self-energies $\Pi^{ }(K;\bmu)$ 
of the neutral Higgs fields $h$;
of the $W$ boson; 
of the top quark; and of the new scalars $H,A$ and $H^\pm$. 
For Standard Model particles the expressions read 
(here $v^{ }_0$ is the {\em tree-level} vacuum expectation value
which can within the 1-loop expressions be approximated as 
$
 v_0^2 \equiv \mu_1^2(\bmu)/\lambda^{ }_1(\bmu) \approx 4 \mW^2 / g_0^2
$):
\ba
 && \hspace*{-3.0cm} \Pi^{ }_h(-i \mh;\bmu) \; = \; 
 \mh^2 \, \delta Z^{ }_{\mu_1^2} 
 \nn 
 & + & 12 h_t^2 A(\mt)
 - 6 \lambda^{ }_1 \, A(\mh)
 + \biggl[ \frac{3(1-D)g_2^2}{2} - 6 \lambda^{ }_1
   \biggr] \, A(\mW)
 \nn & - & 
 2 \lambda^{ }_3 \, A(\mHpm)
 \; - \; 
 \bigl( \lambda^{ }_3 + \lambda^{ }_4 + \lambda^{ }_5 \bigr) \, A(\mH)
 \; - \; 
 \bigl( \lambda^{ }_3 + \lambda^{ }_4 - \lambda^{ }_5 \bigr) \, A(\mA)
 \nn[2mm] & + & 
 3 h_t^2 ( 4 \mt^2 - \mh^2) B(-i\mh;\mt,\mt)
 - 9 \lambda^{ }_1 \mh^2 
 \, B(-i\mh;\mh,\mh)
 \nn
 & + & \Bigl\{  \frac{3g_2^2}{2} \Bigl[ (1-D) \mW^2 + \mh^2 \Bigr]
 - 3 \lambda^{ }_1 \mh^2 \Bigr\}
 \, B(-i\mh;\mW,\mW)
 \nn
 & - & 
 \lambda_3^2 v_0^2 \, B(-i\mh;\mHpm,\mHpm) 
 \; - \; \frac{1}{2} 
 \bigl( \lambda^{ }_3 + \lambda^{ }_4 + \lambda^{ }_5 \bigr)^2
 v_0^2  \, B(-i\mh;\mH,\mH)
 \nn 
 & - & \frac{1}{2}
 \bigl( \lambda^{ }_3 + \lambda^{ }_4 - \lambda^{ }_5 \bigr)^2
 v_0^2  \, B(-i\mh;\mA,\mA)
 \;, \\ 
  && \hspace*{-3.0cm} \Pi^{(T)}_{\rmii{$W$}}(-i \mW,\bmu) \; = \; 
 \mW^2 \, \bigl(
 \delta Z^{ }_{g_2^2}
 - \delta Z^{ }_{\lambda_1} + \delta Z^{ }_{\mu_1^2}  \bigr)
 \nn & + & g_2^2 
 \, \biggl\{
   \biggl( \frac{6 \mt^2}{\mh^2} - \fr32 \biggr) \, A(\mt)
  - \frac{ A(\mh) }{2}
  + \biggl[
       \frac{3(1-D)\mW^2}{2\mh^2} + 2D - 4 
    \biggr] \, A(\mW)
 \nn
 \hspace*{18mm} & + & 
   \biggl[ \fr12 - \frac{\lambda^{ }_3 v_0^2}{2 \mh^2} \biggr] \, A(\mHpm)
 \; + \; 
   \biggl[ 
     \fr14 - \frac{(\lambda^{ }_3 + \lambda^{ }_4 + \lambda^{ }_5)v_0^2}
      {4 \mh^2}
   \biggr]\, A(\mH)
 \nn & + & 
   \biggl[ 
     \fr14 - \frac{(\lambda^{ }_3 + \lambda^{ }_4 - \lambda^{ }_5)v_0^2}
      {4 \mh^2}
   \biggr]\, A(\mA)
 \nn 
 & + & 
   6 \mW^2 B(-i\mW;\mW,\mW) - \mW^2 B(-i\mW;\mW,\mh) 
 \nn & + & 
 \frac{3(\mt^2 - \mW^2)}{2}\, B(-i \mW;0,\mt)
  + \Bigl( \fr32 - 2 \nG \Bigr)\, \mW^2\, B(-i \mW;0,0)
 \nn 
 & + & 
  \bigl( 7 - 4 D \bigr) \, D^{(T)}_{ }(-i\mW;\mW,\mW)  
  - D^{(T)}_{ }(-i\mW;\mW,\mh)
  \nn[2mm] 
  & - & 
   D^{(T)}_{ }(-i\mW;\mHpm,\mH)
  \; - \; 
   D^{(T)}_{ }(-i\mW;\mHpm,\mA)
  \nn[2mm] 
  & + & 
   6 \,
   D^{(T)}_{ }(-i\mW;0,\mt)
  \; + \; 
   \bigl( 8\nG - 6 \bigr) \,
   D^{(T)}_{ }(-i\mW;0,0)
 \biggr\} 
 \;, \\  
 && \hspace*{-3.2cm}
 2 \bigl[ \Sigma^{ }_\rmii{S} (-i\mt;\bmu)
 \,-\, \Sigma^{ }_\rmii{V} (-i\mt;\bmu) \bigr] 
 \; = \;
  \delta Z^{ }_{h_t^2}
 - \delta Z^{ }_{\lambda_1} + \delta Z^{ }_{\mu_1^2} 
 \nn 
 & + & \frac{1}{\mh^2}\, \biggl\{ 12 h_t^2 A(\mt)
 - 6 \lambda^{ }_1 \, A(\mh)
 + \biggl[ \frac{3(1-D)g_2^2}{2} - 6 \lambda^{ }_1
   \biggr] \, A(\mW)
 \nn & - & 
 2 \lambda^{ }_3 \, A(\mHpm)
 \; - \; 
 \bigl( \lambda^{ }_3 + \lambda^{ }_4 + \lambda^{ }_5 \bigr) \, A(\mH)
 \; - \; 
 \bigl( \lambda^{ }_3 + \lambda^{ }_4 - \lambda^{ }_5 \bigr) \, A(\mA)
 \biggr\}
 \nn & + & 
 \frac{8 D g_3^2}{3} \, B(-i\mt;0,\mt) 
 + h_t^2 \Bigl[ B(-i\mt;\mW,\mt) - B(-i\mt;\mh,\mt) \Bigr]
 \nn & + &   
 \frac{8 (D-2) g_3^2 }{3} \, C(-i\mt;0,\mt)
 \nn & + & 
 \frac{(D-2) g_2^2 }{4} \,  
 \Bigl[ 2 C(-i\mt;\mW,0) + C(-i\mt;\mW,\mt)  \Bigr]
 \nn & + & h_t^2 \Bigl[ C(-i\mt;\mh,\mt) + C(-i\mt;\mW,\mt)
 + C(-i\mt;\mW,0) \Bigr]
 \;. 
\ea
For $W$ only the transverse parts play a role, and 
$D^{(T)}_{ }$ is defined by
$D^{ }_{\mu\nu} \equiv D^{(T)}_{ } \delta^{ }_{\mu\nu}
 + \rmO(K^{ }_\mu K^{ }_\nu)$. 
In the case of the top quark the self-energy was expressed as 
$
 \Pi^{ }_t 
 = i\bsl{K} \Sigma^{ }_\rmii{V} 
 + i\bsl{K}\! \gamma^{ }_5 \Sigma^{ }_\rmii{A} 
 + \mt\, \Sigma^{ }_\rmii{S} 
$; bracketing this with on-shell spinors eliminates
the function $ \Sigma^{ }_\rmii{A} (K;\bmu) $.

For the on-shell self-energies of the new scalars we obtain  
(denoting $n^{ }_3 = n^{ }_4 = -n^{ }_5 \equiv 1$)
\ba
 && \hspace*{-3.0cm}
 \Pi^{ }_{\rmii{$H$}}(-i\mH;\bmu) \; = \; 
 \mu_2^2\, \delta Z^{ }_{\mu_2^2}
 + \sum_{i=3,4,5}
 \frac{\lambda_i v_0^2}{2}
 \bigl( \delta Z^{ }_{\lambda^{ }_i} + \delta Z^{ }_{\mu_1^2} - 
 \delta Z^{ }_{\lambda^{ }_1}\bigr)
 \nn & + & 
 \frac{12 (\lambda^{ }_3 + \lambda^{ }_4 + \lambda^{ }_5)\mt^2}{\mh^2}
 \, A(\mt)
 \; - \; (\lambda^{ }_3 + \lambda^{ }_4 + \lambda^{ }_5) \, A(\mh)
 \nn & + & 
 \biggl[
 \frac{3(1-D)(\lambda^{ }_3 + \lambda^{ }_4 + \lambda^{ }_5)\mW^2  }
 {\mh^2}
  + \frac{3(D-2)g_2^2}{4}
  - \lambda^{ }_4 - 2 \lambda^{ }_5  \biggr]\, A(\mW)
 \nn \hspace*{18mm }& + & 
 \biggl[ 3 \lambda^{ }_2
  - \frac{(\lambda^{ }_3 + \lambda^{ }_4 + \lambda^{ }_5)^2 v_0^2}
         {2\mh^2} \biggr] \, A(\mH)
 \nn & + & 
 \biggl[ \lambda^{ }_2 + \frac{g_2^2}{4}
  + \frac{\lambda^{2}_5 v_0^2 - (\lambda^{ }_3 + \lambda^{ }_4)^2 v_0^2}
         {2\mh^2} \biggr] \, A(\mA)
 \nn & + &
 \biggl[ 2 \lambda^{ }_2 + \frac{g_2^2}{2}
  - \frac{\lambda^{ }_3(\lambda^{ }_3 + \lambda^{ }_4 + \lambda^{ }_5) v_0^2}
         {\mh^2} \biggr] \, A(\mHpm)
 \nn & - & 
 (\lambda^{ }_3 + \lambda^{ }_4 + \lambda^{ }_5)^2 v_0^2 
 B(-i\mH;\mh,\mH)
 \nn[2mm] & - & 
 \biggl[ \lambda_5^2 v_0^2 + \frac{(\mW^2 - 2 \mH^2 - 2 \mA^2)g_2^2}{4}
 \biggr]\, B(-i\mH;\mW,\mA)
 \nn & - & 
 \biggl[ \frac{(\lambda^{ }_4 + \lambda^{ }_5)^2 v_0^2}{2} 
 + \frac{(\mW^2 - 2 \mH^2 - 2 \mHpm^2)g_2^2}{2}
 \biggr]\, B(-i\mH;\mW,\mHpm)
 \;, \\ 
 && \hspace*{-3.0cm}
 \Pi^{ }_{\rmii{$A$}}(-i\mA;\bmu) \; = \; 
 \mu_2^2\, \delta Z^{ }_{\mu_2^2}
 + \sum_{i=3,4,5}
 \frac{n^{ }_i \lambda_i v_0^2}{2}
 \bigl( \delta Z^{ }_{\lambda^{ }_i} + \delta Z^{ }_{\mu_1^2} - 
 \delta Z^{ }_{\lambda^{ }_1}\bigr)
 \nn & + & 
 \frac{12 (\lambda^{ }_3 + \lambda^{ }_4 - \lambda^{ }_5)\mt^2}{\mh^2}
 \, A(\mt)
 \; - \; (\lambda^{ }_3 + \lambda^{ }_4 - \lambda^{ }_5) \, A(\mh)
 \nn & + & 
 \biggl[
 \frac{3(1-D)(\lambda^{ }_3 + \lambda^{ }_4 - \lambda^{ }_5) \mW^2  }
 {\mh^2}
  + \frac{3(D-2)g_2^2}{4}
  - \lambda^{ }_4 + 2 \lambda^{ }_5  \biggr]\, A(\mW)
 \nn & + & 
 \biggl[ 3 \lambda^{ }_2
  - \frac{(\lambda^{ }_3 + \lambda^{ }_4 - \lambda^{ }_5)^2 v_0^2}
         {2\mh^2} \biggr] \, A(\mA)
 \nn & + & 
 \biggl[ \lambda^{ }_2 + \frac{g_2^2}{4}
  + \frac{\lambda^{2}_5 v_0^2 - (\lambda^{ }_3 + \lambda^{ }_4)^2 v_0^2}
         {2\mh^2} \biggr] \, A(\mH)
 \nn & + &
 \biggl[ 2 \lambda^{ }_2 + \frac{g_2^2}{2}
  - \frac{\lambda^{ }_3(\lambda^{ }_3 + \lambda^{ }_4 - \lambda^{ }_5) v_0^2}
         {\mh^2} \biggr] \, A(\mHpm)
 \nn & - & 
 (\lambda^{ }_3 + \lambda^{ }_4 - \lambda^{ }_5)^2 v_0^2 
 B(-i\mA;\mh,\mA)
 \nn[2mm] & - & 
 \biggl[ \lambda_5^2 v_0^2 + \frac{(\mW^2 - 2 \mH^2 - 2 \mA^2)g_2^2}{4}
 \biggr]\, B(-i\mA;\mW,\mH)
 \nn & - & 
 \biggl[ \frac{(\lambda^{ }_4 - \lambda^{ }_5)^2 v_0^2}{2} 
 + \frac{(\mW^2 - 2 \mA^2 - 2 \mHpm^2)g_2^2}{2}
 \biggr]\, B(-i\mA;\mW,\mHpm)
 \;, \hspace*{6mm} \\ 
 && \hspace*{-3.0cm}
 \Pi^{ }_{\rmii{$H^{ }_\pm$}}(-i\mHpm;\bmu) \; = \; 
 \mu_2^2\, \delta Z^{ }_{\mu_2^2}
 + 
 \frac{\lambda^{ }_3 v_0^2}{2}
 \bigl( \delta Z^{ }_{\lambda^{ }_3} + \delta Z^{ }_{\mu_1^2} - 
 \delta Z^{ }_{\lambda^{ }_1}\bigr)
 \nn & + & 
 \frac{12 \lambda^{ }_3 \mt^2}{\mh^2}
 \, A(\mt)
 \; - \; \lambda^{ }_3 \, A(\mh)
 \nn & + & 
 \biggl[
 \frac{3(1-D) \lambda^{ }_3 \mW^2 }
 {\mh^2}
  + \frac{3(D-2)g_2^2}{4}
  + \lambda^{ }_4  \biggr]\, A(\mW)
 \nn & + & 
 \biggl[  \lambda^{ }_2 + \frac{g_2^2}{4}
  - \frac{\lambda^{ }_3(\lambda^{ }_3 + \lambda^{ }_4 + \lambda^{ }_5) v_0^2}
         {2\mh^2} \biggr] \, A(\mH)
 \nn & + & 
 \biggl[ \lambda^{ }_2 + \frac{g_2^2}{4}
  - \frac{\lambda^{ }_3(\lambda^{ }_3 + \lambda^{ }_4 - \lambda^{ }_5) v_0^2}
         {2\mh^2} \biggr] \, A(\mA)
 \nn & + &
 \biggl[ 4 \lambda^{ }_2 + \frac{g_2^2}{4}
  - \frac{\lambda^{2}_3 v_0^2}
         {\mh^2} \biggr] \, A(\mHpm)
 \; - \; 
 \lambda^{2}_3 v_0^2 
 B(-i\mHpm;\mh,\mHpm)
 \nn[2mm] & - & 
 \biggl[ \frac{(\lambda^{ }_4 + \lambda^{ }_5)^2 v_0^2}{4} 
 + \frac{(\mW^2 - 2 \mH^2 - 2 \mHpm^2)g_2^2}{4}
 \biggr]\, B(-i\mHpm;\mW,\mH)
 \nn & - & 
 \biggl[ \frac{(\lambda^{ }_4 - \lambda^{ }_5)^2 v_0^2}{4} 
 + \frac{(\mW^2 - 2 \mA^2 - 2 \mHpm^2)g_2^2}{4}
 \biggr]\, B(-i\mHpm;\mW,\mA)
 \nn & - & 
 \biggl[ 
  \frac{(\mW^2 - 4 \mHpm^2)g_2^2}{4}
 \biggr]\, B(-i\mHpm;\mW,\mHpm)
 \;.
\ea

The on-shell self-energies correct tree-level masses, which can be 
expressed in terms of $\msbar$ parameters. For instance, the 
physical Higgs mass squared has the form 
$
 \mh^2 = - 2 \mu_1^2(\bmu) + \re \Pi^{ }_h(-i \mh;\bmu)
$.
Here we defined the pole mass $\mh$ through the projection of the 
complex pole to the real axis.  
The pole mass equation can be inverted to give
\be
 \mu_1^2(\bmu) \quad \stackrel{\bmu \;\approx\; \mZ}{ = } \quad 
 - \frac{\mh^2}{2}
 \, \biggl[ 
   1 - \frac{ \re \Pi^{ }_h(-i \mh;\bmu) }{\mh^2}
 \biggr]
 \;. \la{mu1_renorm}
\ee
Similarly, the other parameters read (always implicitly 
assuming $\bmu \approx \mZ$)
\ba
 \lambda^{ }_1(\bmu) & = & 
 \frac{g_0^2 \mh^2}{8\mW^2}
 \, 
 \biggl[ 1 + \frac{\delta g_2^2(\bmu)}{g_0^2}
 + \frac{\re \Pi_\rmii{$W$}^{(T)}(-i\mW;\bmu)}{\mW^2}
 - \frac{ \re \Pi^{ }_h(-i \mh;\bmu) }{\mh^2}
 \biggr]
 \;, \la{lam1_renorm} \\ 
 h_t^2(\bmu) & = & 
 \frac{g_0^2 \mt^2}{2\mW^2}
 \, 
 \biggl[ 1 + \frac{\delta g_2^2(\bmu)}{g_0^2}
 + \frac{\re \Pi_\rmii{$W$}^{(T)}(-i\mW;\bmu)}{\mW^2}
 \,-\, 2 ( \Sigma^{ }_\rmii{S} - \Sigma^{ }_\rmii{V})(-i\mt;\bmu)
 \biggr]
 \;, \hspace*{7mm} \\ 
 \mu_2^2(\bmu) & = & 
 \mH^2
 \, \biggl[ 
   1 - \frac{ \re \Pi^{ }_\rmii{$H$} (-i \mH;\bmu) }{\mH^2}
 \biggr]
 \nn 
 & - & \frac{2\lambda^{ }_\rmii{$H$}(\bmu)\mW^2}{g_0^2}
 \, \biggl[ 
 1 - \frac{\delta g_2^2(\bmu)}{g_0^2}
 - \frac{\re \Pi_\rmii{$W$}^{(T)}(-i\mW;\bmu)}{\mW^2}
 \biggr]
 \;, \\ 
 \lambda^{ }_3(\bmu) & = & 
 \frac{g_0^2 \mHpm^2}{2\mW^2}
 \, 
 \biggl[ 1 + \frac{\delta g_2^2(\bmu)}{g_0^2}
 + \frac{\re \Pi_\rmii{$W$}^{(T)}(-i\mW;\bmu)}{\mW^2}
 - \frac{ \re \Pi^{ }_\rmii{$H_\pm$}(-i \mHpm;\bmu) }{\mHpm^2}
 \biggr]
 \nn & - & 
  \frac{g_0^2 \mH^2}{2\mW^2}
 \, 
 \biggl[ 1 + \frac{\delta g_2^2(\bmu)}{g_0^2}
 + \frac{\re \Pi_\rmii{$W$}^{(T)}(-i\mW;\bmu)}{\mW^2}
 - \frac{ \re \Pi^{ }_\rmii{$H$}(-i \mH;\bmu) }{\mH^2}
 \biggr]
 \nn & + & 
 \lambda^{ }_\rmii{$H$}(\bmu)
 \, \biggl[ 
 1 - \frac{\delta g_2^2(\bmu)}{g_0^2}
 - \frac{\re \Pi_\rmii{$W$}^{(T)}(-i\mW;\bmu)}{\mW^2}
 \biggr]
 \;, \\ 
 \lambda^{ }_4(\bmu) & = & 
 \frac{g_0^2 \mH^2}{4\mW^2}
 \, 
 \biggl[ 1 + \frac{\delta g_2^2(\bmu)}{g_0^2}
 + \frac{\re \Pi_\rmii{$W$}^{(T)}(-i\mW;\bmu)}{\mW^2}
 - \frac{ \re \Pi^{ }_\rmii{$H$}(-i \mH;\bmu) }{\mH^2}
 \biggr]
 \nn & + & 
 \frac{g_0^2 \mA^2}{4\mW^2}
 \, 
 \biggl[ 1 + \frac{\delta g_2^2(\bmu)}{g_0^2}
 + \frac{\re \Pi_\rmii{$W$}^{(T)}(-i\mW;\bmu)}{\mW^2}
 - \frac{ \re \Pi^{ }_\rmii{$A$}(-i \mA;\bmu) }{\mA^2}
 \biggr]
 \nn & - & 
 \frac{g_0^2 \mHpm^2}{2\mW^2}
 \, 
 \biggl[ 1 + \frac{\delta g_2^2(\bmu)}{g_0^2}
 + \frac{\re \Pi_\rmii{$W$}^{(T)}(-i\mW;\bmu)}{\mW^2}
 - \frac{ \re \Pi^{ }_\rmii{$H_\pm$}(-i \mHpm;\bmu) }{\mHpm^2}
 \biggr]
 \nn & + & 
 \frac{\lambda^{ }_\rmii{$H$}(\bmu)}{2}
 \, \biggl[ 
 \frac{\delta g_2^2(\bmu)}{g_0^2}
 + \frac{\re \Pi_\rmii{$W$}^{(T)}(-i\mW;\bmu)}{\mW^2}
 \biggr]
 \;, \\ 
 \lambda^{ }_5(\bmu) & = & 
 \frac{g_0^2 \mH^2}{4\mW^2}
 \, 
 \biggl[ 1 + \frac{\delta g_2^2(\bmu)}{g_0^2}
 + \frac{\re \Pi_\rmii{$W$}^{(T)}(-i\mW;\bmu)}{\mW^2}
 - \frac{ \re \Pi^{ }_\rmii{$H$}(-i \mH;\bmu) }{\mH^2}
 \biggr]
 \nn & - & 
 \frac{g_0^2 \mA^2}{4\mW^2}
 \, 
 \biggl[ 1 + \frac{\delta g_2^2(\bmu)}{g_0^2}
 + \frac{\re \Pi_\rmii{$W$}^{(T)}(-i\mW;\bmu)}{\mW^2}
 - \frac{ \re \Pi^{ }_\rmii{$A$}(-i \mA;\bmu) }{\mA^2}
 \biggr]
 \nn & + & 
 \frac{\lambda^{ }_\rmii{$H$}(\bmu)}{2}
 \, \biggl[ 
 \frac{\delta g_2^2(\bmu)}{g_0^2}
 + \frac{\re \Pi_\rmii{$W$}^{(T)}(-i\mW;\bmu)}{\mW^2}
 \biggr]
 \;, \la{lam5_renorm}
\ea
where $\delta g_2^2(\bmu) \equiv g_2^2(\bmu) - g_0^2$
is from \eq\nr{g2}. 
Following conventions
in the literature, $\lambda^{ }_2(\mZ)$ and 
$\lambda^{ }_\rmii{$H$} (\mZ) \equiv 
\lambda^{ }_3(\mZ) + \lambda^{ }_4(\mZ) + \lambda^{ }_5(\mZ)$
are used directly as input parameters. 

For approximate estimates, including only the large
effects from $\lambda^{2}_3$, 
$(\lambda^{ }_3 + \lambda^{ }_4 \pm \lambda^{ }_5)^2$ and $h_t^4$, 
\eqs\nr{mu1_renorm} and \nr{lam1_renorm} can be simplified into  
\ba
 \mu_1^2(\bmu^{ }) & \simeq & 
 - \frac{\mh^2}{2}  
 \; +  \; \frac{1}{32\pi^2}
 \biggl[ 
    2 \lambda^{ }_3
    \biggl(\mHpm^2 + \mu_2^2  \ln \frac{\bmu^2_{ }}{\mHpm^2} \biggr)
  - 12 h_t^2 \mt^2
 \\ & + & 
    (\lambda^{ }_3 + \lambda^{ }_4 + \lambda^{ }_5)
    \biggl(\mH^2 + \mu_2^2  \ln \frac{\bmu^2_{ }}{\mH^2} \biggr)
  +
    (\lambda^{ }_3 + \lambda^{ }_4 - \lambda^{ }_5)
    \biggl(\mA^2 + \mu_2^2  \ln \frac{\bmu^2_{ }}{\mA^2} \biggr)
 \biggr]
 \;, \hspace*{8mm} \nn 
 \lambda^{ }_1(\bmu^{ }) & \simeq & 
 \frac{g_2^2 \mh^2}{8 \mW^2}
 + \frac{1}{64\pi^2}
 \biggl[ 
   2 \lambda_3^2 \ln \frac{\bmu^2_{ }}{\mHpm^2}
  - 12 h_t^4 \ln \frac{\bmu^2_{ }}{\mt^2}
 \la{lam1_rough} \\ & + & 
  (\lambda^{ }_3 + \lambda^{ }_4 + \lambda^{ }_5)^2 
  \ln \frac{\bmu^2_{ }}{\mH^2} 
  +
  (\lambda^{ }_3 + \lambda^{ }_4 - \lambda^{ }_5)^2 
  \ln \frac{\bmu^2_{ }}{\mA^2} 
 \biggr]
 \;. \nonumber 
\ea
Apart from directly approximating 
\eqs\nr{mu1_renorm} and \nr{lam1_renorm}, these expressions
can also be derived from the ``naive'' procedure of 
minimizing the effective potential $V^{ }_0 + V^{ }_1$, 
and tuning $\mu_1^2(\mZ)$ and $\lambda^{ }_1(\mZ)$ so that 
the location of the minimum is at
$v_\rmi{min}^2 \simeq 4 \mW^2 / g_2^2$ 
and the second derivative at the minimum is 
$(V^{ }_0 + V^{ }_1)''(v^{ }_\rmi{min}) \simeq \mh^2$. 
This naive procedure can easily be implemented numerically
and then also applied to the 2-loop potential at zero temperature. 

%
\subsection{Practical procedure}
\la{ss:practice}

A problem with the 1-loop expressions listed in 
appendix~\ref{ss:couplingsmZ}
is that if the couplings $\lambda^{ }_3, \lambda^{ }_4$ and $\lambda^{ }_5$
are first determined at tree level, and these values are 
subsequently inserted into the 1-loop corrections, as given 
in \eqs\nr{mu1_renorm}--\nr{lam5_renorm}, then the corrections are
in many cases of order 100\%; for instance, $\lambda^{ }_1(\mZ)$ can
be driven to a negative value. 
If corrections are of order 100\%, 
there is no reason to trust the results. The problem
can be somewhat ``regulated'' by solving 
\eqs\nr{mu1_renorm}--\nr{lam5_renorm} ``self-consistently'', 
i.e.\ by requiring that the couplings have the same values on 
both sides of the equations. In general, this reduces the 
magnitude of the largest coupling $\lambda^{ }_3$, whereby
the corrections remain below 100\%. A further ``resummation'' 
can be implemented by determining $\mu_1^2(\mZ)$ and 
$\lambda^{ }_1(\mZ)$ {\em \`a la} Coleman-Weinberg, 
as outlined below \eq\nr{lam1_rough}. An advantage of this 
procedure is that 2-loop corrections can be partially included
into $\mu_1^2(\mZ)$ and $\lambda^{ }_1(\mZ)$. The values listed
in table~\ref{table:couplingsmZ} have been obtained by determining
$\mu_1^2(\mZ)$ and $\lambda^{ }_1(\mZ)$ from the effective potential
and the other parameters from the 1-loop pole mass relations
in \se\ref{ss:couplingsmZ}. However, 
we have also tested the procedure where 1-loop pole mass
relations are used for all the couplings. This changes the values of
$\mu_1^2(\mZ)$ and $\lambda^{ }_1(\mZ)$ given in 
table~\ref{table:couplingsmZ} by up to $\sim 20$\% for BM2,
and only a few \% for BM1 and BM3, however 
our conclusions concerning thermal effects remain unchanged in all cases.

%
\subsection{Counterterms and renormalization group equations}
\la{ss:RG}

Finally, let us list the counter\-terms needed in our analysis. 
The notation for them was defined in \se\ref{ss:cts}. 
We stress that the same 
counter\-terms appear both in the vacuum renormalization
computations of the current section and in  
the thermal 2-loop effective potential
given in appendix~B. 
The results 
agree with ref.~\cite{pole} for the counterterms that can be 
found there.  The complete list reads
\ba
 \delta Z^{ }_{\mu^{2}_1} & = & 
 \frac{1}{(4\pi)^2\epsilon}
 \biggl[ 
  3 h_t^2 - \frac{9 g_2^2}{4} 
 + 6 \lambda^{ }_1 
 + \frac{\mu_2^2}{\mu_1^2}
 \bigl( 2 \lambda^{ }_3 + \lambda^{ }_4\bigr)
 \biggr]
 \;, \\ 
 \delta Z^{ }_{\mu^{2}_2} & = & 
 \frac{1}{(4\pi)^2\epsilon}
 \biggl[ 
  - \frac{9 g_2^2}{4} 
 + 6 \lambda^{ }_2 
 + \frac{\mu_1^2}{\mu_2^2}
 \bigl( 2 \lambda^{ }_3 + \lambda^{ }_4\bigr)
 \biggr]
 \;, \\ 
 \delta Z^{ }_{\lambda^{ }_1} & = & 
 \frac{1}{(4\pi)^2\epsilon}
 \biggl[ 
  6 h_t^2 - \frac{9 g_2^2}{2} +\frac{9 g_2^4}{16 \lambda^{ }_1}
 - \frac{3 h_t^4}{\lambda^{ }_1} + 12 \lambda^{ }_1
 + \frac{
 2 \lambda^{2}_3 + 2 \lambda^{ }_3 \lambda^{ }_4 
 + \lambda_4^2 + \lambda_5^2 
 }{2 \lambda^{ }_1}
 \biggr]
 \;, \\ 
 \delta Z^{ }_{\lambda^{ }_2} & = & 
 \frac{1}{(4\pi)^2\epsilon}
 \biggl[ 
 - \frac{9 g_2^2}{2} +\frac{9 g_2^4}{16 \lambda^{ }_2}
 + 12 \lambda^{ }_2
 + \frac{
 2 \lambda^{2}_3 + 2 \lambda^{ }_3 \lambda^{ }_4 
 + \lambda_4^2 + \lambda_5^2 
 }{2 \lambda^{ }_2}
 \biggr]
 \;, \\ 
 \delta Z^{ }_{\lambda^{ }_3} & = & 
 \frac{1}{(4\pi)^2\epsilon}
 \biggl[ 
  3 h_t^2 - \frac{9 g_2^2}{2} +\frac{9 g_2^4}{8\lambda^{ }_3}
 + 6 (\lambda^{ }_1 + \lambda^{ }_2)
 + 2 \lambda^{ }_3 + \frac{
 2 (\lambda^{ }_1 + \lambda^{ }_2) \lambda^{ }_4 
 + \lambda_4^2 + \lambda_5^2 
 }{\lambda^{ }_3}
 \biggr]
 \;, \hspace*{1cm} \\ 
 \delta Z^{ }_{\lambda^{ }_4} & = & 
 \frac{1}{(4\pi)^2\epsilon}
 \biggl[ 
  3 h_t^2 - \frac{9 g_2^2}{2} 
 + 2 (\lambda^{ }_1 + \lambda^{ }_2)
 + 4 \lambda^{ }_3 + 2 \lambda^{ }_4 + \frac{
 4 \lambda_5^2 
 }{\lambda^{ }_4}
 \biggr]
 \;, \\ 
 \delta Z^{ }_{\lambda^{ }_5} & = & 
 \frac{1}{(4\pi)^2\epsilon}
 \biggl[ 
  3 h_t^2 - \frac{9 g_2^2}{2} + 2 (\lambda^{ }_1 + \lambda^{ }_2)
 + 4 \lambda^{ }_3 + 6 \lambda^{ }_4
 \biggr]
 \;, \\ 
 \delta Z^{ }_{g^2_{2}} & = & 
 \frac{g_2^2}{(4\pi)^2\epsilon}
 \biggl[ \frac{4\nG}{3} - 7 \biggr]
 \;, \\ 
 \delta Z^{ }_{g^2_{3}} & = & 
 \frac{g_3^2}{(4\pi)^2\epsilon}
 \biggl[ \frac{4\nG}{3} - 11 \biggr]
 \;, \\ 
 \delta Z^{ }_{h^{2}_t} & = & 
 \frac{1}{(4\pi)^2\epsilon}
 \biggl[
  \frac{9 h_t^2}{2} - \frac{9 g_2^2}{4} - 8 g_3^2 
 \biggr]
 \;, \\ 
 \delta Z^{ }_{\phi} & = & \frac{1}{(4\pi)^2\epsilon}
 \biggl[
  - 3h_t^2  + 3g_2^2
 \biggr]
 \;, \la{Z_phi}\\ 
 \delta Z^{ }_{v} & = & \frac{1}{(4\pi)^2\epsilon}
 \biggl[
 3h_t^2 + \frac{5g_2^2}{2} 
 \biggr]
 \;, \la{Z_v}\\ 
 \delta Z^{ }_{\chi} & = & \frac{1}{(4\pi)^2\epsilon}
 \biggl[ 
  \frac{3 g_2^2}{2}
 \biggr]
 \;, \\ 
 \delta Z^{ }_{A} & = & 
 \delta Z^{ }_\xi \; = \; \frac{g_2^2}{(4\pi)^2\epsilon}
 \biggl[ 3 - \frac{4\nG}{3} \biggr]
 \;. \la{cts}
\ea
The counterterm $\delta Z^{ }_{\chi}$, appearing in \eq\nr{s}, only
contributes to thermal effects which are formally of higher order than
the accuracy of the computation. 

As usual, the counterterms fix the renormalization group equations as
\be
 \bmu \frac{{\rm d}\lambda^{ }_i}{{\rm d}\bmu}
 \; = \; 
 2 \lambda^{ }_i \, \epsilon\, \delta Z^{ }_{\lambda_i} + \rmO(\lambda_i^3)
 \;, \la{RG}
\ee
and similarly for the other couplings. 

%

\end{document}